\documentclass[aps,pra,reprint,superscriptaddress,twocolumn,showkeys,showpacs,amsmath,amssymb,longbibliography,floatfix]{revtex4-2}

\usepackage{amssymb,amsmath,amsthm}
\usepackage{mathtools}
\usepackage{graphicx}
\usepackage[english]{babel}

\usepackage{silence}
\usepackage{float}

\usepackage[svgnames]{xcolor}
\usepackage{blindtext}
\usepackage{bm}
\usepackage{comment}
\usepackage{etoolbox}
\usepackage{lipsum}
\usepackage{braket}
\usepackage{relsize}
\usepackage[symbol]{footmisc}

\usepackage[colorlinks,linkcolor=blue,citecolor=blue,urlcolor=blue]{hyperref}

\begin{document}

\title{ \texorpdfstring{$N$-dimensional discrete Fourier transform via bosonic Hamiltonian}{N-dimensional discrete Fourier transform via bosonic Hamiltonian} }

\author{Edgar Barriga}
\affiliation{Millennium Institute for Research in Optics (MIRO), Chile}
\affiliation{Departamento de F\'{\i}sica, Facultad de Ciencias, Universidad de Chile, Santiago, Chile}

\author{Camila Mu\~noz}
\affiliation{Departamento de F\'isica, Universidad de Concepci\'on, 160-C Concepci\'on, Chile}

\author{Alejandro Mu\~noz}
\affiliation{Departamento de F\'{\i}sica,  Facultad de Ciencias F\'isicas y Matem\'aticas, Universidad de Chile, Santiago, Chile}
\affiliation{Millennium Institute for Research in Optics (MIRO), Chile}

\author{Santiago Rojas-Rojas}

\affiliation{Millennium Institute for Research in Optics (MIRO), Chile}

\author{Pablo Solano}
\affiliation{Departamento de F\'isica, Universidad de Concepci\'on, 160-C Concepci\'on, Chile}

\author{Carla Hermann-Avigliano}
\affiliation{Departamento de F\'{\i}sica,  Facultad de Ciencias F\'isicas y Matem\'aticas, Universidad de Chile, Santiago, Chile}
\affiliation{Millennium Institute for Research in Optics (MIRO), Chile}

\author{Aldo Delgado}
\affiliation{Departamento de F\'isica, Universidad de Concepci\'on, 160-C Concepci\'on, Chile}
\affiliation{Millennium Institute for Research in Optics (MIRO), Chile}

\begin{abstract}

The discrete Fourier transform (DFT) underpins many classical algorithms and is a fundamental unitary operator for quantum information processing. Implementing the $N$-dimensional DFT in photonic integrated circuits (PICs) is limited by the cascades of Mach-Zehnder interferometers that current architectures require. Here we propose bosonic Hamiltonians that realize the $N$-dimensional DFT through a single stage of multimode evolution, complemented only by phase shifters before and after the interaction region, in a geometry suited to 3D waveguides. Modeling the system as a graph, where edges correspond to couplings and the vertices are the waveguides, we obtain analytical solutions for complete graphs up to $N=6$ and numerical solutions up to $N=31$. For non-complete graphs, different propagation constants are required in the Hamiltonian. We report all solutions for $N\leq 8$, partial exploration for $N=9$, and selected cases for $N=10$, together with three conjectures that guide the numerical search for $N \geq 11$. These configurations circumvent the vanishing evanescent coupling strength imposed by the waveguide separation, and a closed-form sensitivity criterion selects those that are admissible as a waveguide layout and least sensitive to fabrication error. We also uncover the missing non-affine parameters of the $6$-dimensional DFT, and show that the scaling law for implementing the $N$-dimensional DFT with our building blocks is $\mathcal{O}(N\log\log{N})$. This allows assembling the $2520$-dimensional DFT with only $2625$ interferometers, in contrast to the $\approx 3\times 10^6$ of Reck and Clements architectures.

\end{abstract}

\maketitle

\section{Introduction}

The discrete Fourier transform (DFT), a linear transformation that relates spatial or temporal data encoded in a function to its frequency-domain counterpart, is of great importance in many branches of science and engineering~\cite{Briggs1995}. The importance of this operator extend beyond classical computing, and Shor's algorithm~\cite{Shor1994} is a great example where the quantum Fourier transform (QFT) is a key element and in this case allows achieving polynomial time for factorizing large integers. The QFT finds applications in many other quantum algorithms and subroutines~\cite{Kitaev1995,RuizPerez2017,Garcia_Molina2022,Jain2024}, for this reason the optimization of quantum gates used in its implementation is under constant development and improvements~\cite{Nam2020,Park2023}.

Another paradigm for implementing the DFT matrix as a unitary operator acting in a quantum state is the quantum Fourier transform interferometer (QFTI)~\cite{Su2017}, where the waveguide mesh physically defines the unitary evolution operator that maps directly onto the DFT matrix. 
The QFTI plays an important role in the study of multiphoton interference that allows super-sensitivity for single and multiparameter estimation~\cite{You2017}, create number-path entanglement for quantum metrology~\cite{Motes2015}, coherently map photonic states into a qubit register for optical imaging~\cite{Mokeev2025}, a fundamental component of a universal quantum sorter~\cite{Ionicioiu2016}, the transfer matrix for a programmable universal multiport array~\cite{Pereira2025} and being the interferometer to accomplish the certification required in boson sampling by virtue of the suppression law~\cite{tichy2014}, to name a few.

To realize the QFTI, photonic integrated circuits (PICs) are a reliable platform as these devices can implement any unitary operator. This characteristic is achieved through Reck~\cite{Reck1994} and Clements~\cite{Clements2016} architectures, where Mach-Zehnder interferometers (MZIs) are a building block that allow programmability by means of adjustable phase shifters. However, the number of MZIs in these architectures scales as $\mathcal{O}(N^2)$, where $N$ is the number of spatial ports or modes (waveguides). The case of $N=2^p$ was optimized by Barak~\textit{et al.}~\cite{Barak2007} by mapping the Cooley-Tukey algorithm in circuit design, reducing spatial complexity~\cite{Li2025} from $\mathcal{O}(N^2)$ to $\mathcal{O}(N\log{N})$. This new design was implemented in the laboratory by Crespi \textit{et al.}~\cite{Crespi2016} and relied on femtosecond direct laser writing, allowing the fabrication of the $8$-dimensional QFTI in a 3D structure.

The Barak design considerably reduced the number of stages of interaction to implement the QFTI. Despite this progress, the scaling $\mathcal{O}(N\log{N})$ of the circuit is still far from applications beyond the proof-of-principle for the QFTI. An important case in this regard is boson sampling, where to prove genuine interference and demonstrate quantum computational advantage, we need to work with at least $50$ single photons evolving in an interferometer of $N=2500$ modes capable of implementing the QFTI~\cite{tichy2014}.

In this work, we study the evolution of coupled bosonic spatial modes analytically and numerically and present a systematization in the search for geometrical configurations of couplings that can perform the $N$-dimensional QFTI in a single stage of interaction, complemented only by phase shifters before and after the interaction region. Two branches of QFTI design are born from our solutions: regular polygon- and graph-based models. In the former model, we find the equations that govern what couplings will lead to the QFTI, and our calculations strongly suggest that the evolution operator can always converge to a CHM; in the latter, unlike the previous case, different propagation constants must be included in the Hamiltonian.

This paper is organized as follows.
In Sec.~\ref{subsec:2D_3D} we give a mathematical overview of the $N$-dimensional QFTI and briefly review the known PICs where it has been implemented.
In Sec.~\ref{sec:polygonal_model} we study a system of waveguides with all-to-all coupling and present the equations to describe it.
This model suffers limitations with long-rage interactions due to the exponential decay of the evanescent coupling, and 
in Sec.~\ref{sec:new_DFT} we address this problem providing new geometrical configurations that do not require interactions between all neighbors. In Sec.~\ref{sec:numerical_CHM} we introduce the numerical details for obtaining the QFTI in these new configurations and in Sec.~\ref{sec:prime_and_affine} we present the solutions for $N=4,\ 5,\ 7,\ 8,\ 9$ and in Sec.\ref{sec:qfti_N=6} for $N=6$ and some cases for $N=10$. In Sec.~\ref{sec:conjectures} we summarize all our exploration of coupling configurations where we obtained the QFTI and postulate three conjectures to guide the numerical efforts. In Sec.~\ref{sec:tolerance} we establish a framework to analyze the experimental feasibility of the solutions we find in Secs.~\ref{Sec:complete_graph} and~\ref{sec:new_DFT}. In Sec.~\ref{sec:FFT} we extend the results of Barak \textit{et al}.~\cite{Barak2007} by including the Good-Thomas algorithm for the design of the $N$-dimensional QFTI with the new building blocks found in Sec.~\ref{sec:new_DFT}.

\section{\label{Sec:complete_graph} All-to-all coupling}

In this section, we present general equations for describing the couplings in a multi-arm interferometer in which every waveguide interacts with all its neighbors. We recap the formalism of quantum optics behind the generalized Mach-Zehnder interferometers and then introduce the polygonal model in an inductive way. We also determine the experimental limitation of the system.

\subsection{\label{subsec:2D_3D} QFTI in 2D and 3D waveguide architectures}

The interaction between bosonic modes $a_l^\dagger$ in a waveguide array can be described, in the second quantization, through the following Hamiltonian 
\begin{align}\label{eq:Hamiltonian}
    \mathcal{\hat H} &=\hbar\sum\limits_{l} \beta_l a_l^\dagger a_l +\hbar\sum\limits_{\langle j,k\rangle} \left(C_{j k}a_j^\dagger a_k+C_{jk}^\ast a_k^\dagger a_j\right)\,,
\end{align}
where $\beta_l$ is the propagation constant in the $l$-th waveguide and $C_{jk}$ represents the evanescent coupling between the $j$-th and $k$-th modes. The evolution of the annihilation and creation operators is given by a unitary operator $U$ that is obtained as the exponential of the coupling matrix $\mathcal{C}$ in the single particle representation \cite{Rojas-Rojas2024}, that is,
\begin{equation}
U=e^{-i\mathcal{C}z}, 
\label{eq:U=exp_C}
\end{equation}
where $z$ is the interaction or propagation length, and
\begin{equation}
\vec{a}'=U\vec{a},
\end{equation}
with $\vec{a}'$ ($\vec{a}$) a column vector containing the output (input) annihilation modes. For the sake of clarity, hereafter $z$ will be implicitly included in the coefficients $\beta_l$ and $C_{jk}$.

The $N$-dimensional DFT is given by
\begingroup
\setlength\arraycolsep{1.85pt}
\begin{align}\label{eq:dft}
    F_N \!\!\; &= \!\!\;\frac{1}{\sqrt{N}}\!
    \begin{pmatrix}
        1 & 1 & 1 &\cdots & 1\\
        1 & \omega & \omega^2 &\cdots & \omega^{N-1}\\
        1 & \omega^2 & \omega^4 & \cdots& \omega^{2(N-1)} \\
        1 & \omega^3 & \omega^6 & \cdots & \omega^{3(N-1)}\\
        \vdots & \vdots & \vdots &\ddots & \vdots \\
        1 & \omega^{N-1} & \omega^{(N-1)2} & \cdots & \omega^{(N-1)(N-1)}
    \end{pmatrix},
\end{align}
\endgroup
where $\omega=e^{-2i\pi/N}$.

From Eq.~\eqref{eq:U=exp_C} one may be tempted to calculate the matrix logarithm of Eq.~\eqref{eq:dft}, which can be done analytically~\cite{Aristidou2007}. Nevertheless, this analytical solution cannot be implemented in the circuit design as this will involve negative propagation constants. A physically acceptable realization of a QFTI can be obtained by supplementing the unitary evolution of Eq.~\eqref{eq:U=exp_C} with phase-shifters, that is, with diagonal matrices $\Phi^{\rm in}$ and $\Phi^{\rm out}$, before and after the free evolution.  

In the case of two waveguides, the unitary transformation can be easily obtained
\begin{align}
    U_{2}=\exp{\begin{pmatrix}
        0 & -i\theta\\
        -i\theta & 0
    \end{pmatrix}}=\begin{pmatrix}
        \cos{\theta} & -i\sin{\theta}\\
        -i\sin{\theta} & \cos{\theta}
    \end{pmatrix},
\end{align}
where $\theta$ is the coupling coefficient. When the coupling is $\theta = \pi/4$ and the input and output phase-shifter matrices are 
\begin{equation}
\Phi^{\rm out}=\Phi^{\rm in}=\begin{pmatrix}
        1 & 0\\
        0 & e^{i\pi/2}
    \end{pmatrix},
\end{equation}
we obtain the $2$-dimensional QFTI, or equivalently, the Hadamard gate, as
\begin{equation}
\Phi^{\rm out}U_2\Phi^{\rm in}=\frac{1}{\sqrt{2}}
    \begin{pmatrix}
        1 & \phantom{-}1\\
        1 & -1
    \end{pmatrix}.
\end{equation}
From this result, we can see that even in the most simple case, we cannot obtain the DFT directly from Eq.~\eqref{eq:U=exp_C}. Notice that the factor $1/\sqrt{N}$ for $U=F_N$~\eqref{eq:dft} is not given by an arbitrary choice like in the classical DFT, where a factor $1/N$ can be put in the inverse operation, and hence the modulus of the elements in Eq.~\eqref{eq:dft} must be $1/\sqrt{N}$. We will refer to this as the balancing or normalization condition.

The $3$-dimensional QFTI was introduced by \.{Z}ukowski \textit{et al}.~\cite{zukowski1997} in the context of multi-port beamsplitters and was called tritter. Its implementation was based on the pyramid-like design introduced by Reck \textit{et al}.~\cite{Reck1994}. Later, Spagnolo \textit{et al}.~\cite{Spagnolo2013} presented a new realization of the tritter based on 3D integrated optical waveguides. This allowed a single-step evolution realization of the $3$-dimensional DFT. They solved the Heisenberg equation considering the same coupling for all interactions (balanced multi-port) and found the evolution operator to be given by
\begingroup
\setlength\arraycolsep{2pt}
\begin{align}\label{eq:U_3x3}
    U_{3} \!\!\; &= \frac{1}{3}\!
    \begin{pmatrix}
    e^{-2i\theta}+2e^{i\theta} & e^{2i\theta}-e^{i\theta} & e^{2i\theta}-e^{i\theta}\\
    e^{2i\theta}-e^{i\theta} & e^{-2i\theta}+2e^{i\theta} & e^{2i\theta}-e^{i\theta} \\
    e^{2i\theta}-e^{i\theta} & e^{2i\theta}-e^{i\theta} & e^{-2i\theta}+2e^{i\theta}
    \end{pmatrix},
\end{align}
\endgroup
where the balancing to $1/\sqrt{3}$ is achieved when $\theta =2\pi/9$. Selecting phase-shifter matrices as
\begin{equation}
\Phi^{\rm in}=\Phi^{\rm out}=\begin{pmatrix}
    1 & 0 & 0\\
    0 & e^{i 4\pi/3} & 0\\
    0 & 0 & e^{i 4\pi/3}
    \end{pmatrix},    
\end{equation}
the $3$-dimensional DFT is generated as
\begin{align}
    \Phi^{\rm out} U_{3}\Phi^{\rm in}&=\frac{1}{\sqrt{3}}
    \begin{pmatrix}
    1 & 1 &  1\\
    1 & e^{i2\pi/3} & e^{i4\pi/3} \\
    1 & e^{i4\pi/3} & e^{i8\pi/3}
    \end{pmatrix}.
\end{align}

On the other hand, the transformation matrix presented in Eq.~\eqref{eq:U_3x3} can be obtained straightforwardly by means of Eq.~\eqref{eq:U=exp_C} with the coupling matrix
\begin{align}
    \mathcal{C} &= \begin{pmatrix}
        0 & \theta & \theta\\
        \theta & 0 & \theta\\
        \theta & \theta & 0
    \end{pmatrix},
\end{align}
in which case all modes have the same coupling constant.

When we consider the realization of the $4$-dimensional QFTI by means of a symmetric multi-port of four modes, known as the quarter~\cite{zukowski1997}, a new parameter $\phi$ appears in the transformation matrix, which is not fixed by balancing to $1/2$. In fact, we have~\cite{Munoz2024}
\begin{align}\label{eq:Fourier_family_4x4}
    U_{4} &=\frac{1}{2}
    \begin{pmatrix}
        1 & \phantom{-}1 & \phantom{-}1 & \phantom{-}1\\
        1 & \phantom{-}e^{i \phi} & -1 & -e^{i\phi}\\
        1 & -1 & \phantom{-}1 & -1\\
        1 & -e^{i \phi} & -1 & \phantom{-}e^{i \phi}
    \end{pmatrix}.
\end{align}

This phase has been mentioned in experimental works~\cite{Peruzzo2011,Spagnolo2012,Carine2020}, however, among all variables that define how photons evolve in this multi-port interferometer, the origin of the phase $\phi$ remains undisclosed. This phase $\phi$ appears when we calculate Eq.~\eqref{eq:U=exp_C}. To appreciate this, let us consider the coupling in the quarter to be $\theta$ at the edges and $\varphi$ on the diagonals (see Fig.~\ref{fig:polygon23456}). From this specific geometry, the coupling matrix reads
\begin{align}
    \mathcal{C} &= 
    \begin{pmatrix}
        0 & \theta & \varphi & \theta \\
        \theta & 0 & \theta & \varphi \\
        \varphi & \theta & 0 & \theta \\
        \theta & \varphi & \theta & 0
    \end{pmatrix},
\end{align}
yielding the following transformation matrix~\cite{Munoz2024}
\begin{widetext}
\begin{align}\label{eq:U_4x4}
    U_{4} &= \frac{e^{-i(2\theta+\varphi)}}{4}
    \begin{pmatrix}
        1+2e^{2i(\theta+\varphi)}+e^{4i\theta} & 1-e^{4i\theta} & 1-2e^{2i(\theta+\varphi)}+e^{4i\theta} & 1-e^{4i\theta} \\
        1-e^{4i\theta} & 1+2e^{2i(\theta+\varphi)}+e^{4i\theta} & 1-e^{4i\theta} & 1-2e^{2i(\theta+\varphi)}+e^{4i\theta} \\
        1-2e^{2i(\theta+\varphi)}+e^{4i\theta} & 1-e^{4i\theta} & 1+2e^{2i(\theta+\varphi)}+e^{4i\theta} & 1-e^{4i\theta} \\
        1-e^{4i\theta} & 1-2e^{2i(\theta+\varphi)}+e^{4i\theta} & 1-e^{4i\theta} &  1+2e^{2i(\theta+\varphi)}+e^{4i\theta}
    \end{pmatrix}.
\end{align}
\end{widetext}
Balancing the moduli in Eq.~\eqref{eq:U_4x4}, implies that
\begin{subequations}\label{eq:4x4_normalization}
    \begin{align}
    1 &=\left|\sin{\left(2\theta\right)}\right|, \\
    0 &= \left|\cos{\left(2\theta \right)}\sin{\left(2\varphi\right)}\right|,
\end{align}
\end{subequations}
thus, only $\theta$ is relevant to attain normalization, and its value must clearly be $\theta =\pi/4+k\pi$, $k\in\mathbb{Z}$. As next step in the search for the $4$-dimensional QFTI, we need to make the coefficients of the first row and the first column equal to~1. This can be done with the phase-shifter matrices 
\begin{subequations}
    \begin{align}
        \Phi^{\rm in} &= {\rm diag}\left(e^{-i\varphi},ie^{i\varphi}, -e^{-i\varphi}, e^{i\varphi}\right),\\
        \Phi^{\rm out} &={\rm diag}\left(1, ie^{i2\varphi},-1,ie^{i2\varphi} \right),
    \end{align}
\end{subequations}
leaving us with the transformation
\begin{align}\label{eq:DFT4_phases}
    \Phi^{\rm out} U_{4}\Phi^{\rm in} &=\frac{1}{2}
    \begin{pmatrix}
        1 & \phantom{-}1 & \phantom{-}1 & \phantom{-}1\\
        1 & -e^{i4\varphi} & -1 & \phantom{-}e^{i4\varphi} \\
        1 & -1 & \phantom{-}1 & -1\\
        1 & \phantom{-}e^{i4\varphi} & -1 & -e^{i4\varphi}
    \end{pmatrix}.
\end{align}
By comparing Eq.~\eqref{eq:DFT4_phases} with Eq.~\eqref{eq:Fourier_family_4x4} we can see that the phase $\phi$ mentioned in Eq.~\eqref{eq:Fourier_family_4x4} corresponds to the diagonal coupling $\varphi$ in the quarter array (see Fig.~\ref{fig:polygon23456}). When $\theta = \pi/4$, $\varphi=\pi/8$ and after permuting the second with the fourth column in Eq.~\eqref{eq:DFT4_phases}, we finally obtain the $4$-dimensional DFT. In this context, permutations of columns (rows) mean that waveguides must be swapped before (after) the interaction occurs or relabel the waveguide indices.

\subsection{\label{sec:polygonal_model} Polygonal model}

In this subsection, we analytically solve the interactions for a pentagonal arrangement of waveguides to realize the $5$-dimensional QFTI and then generalize the case of waveguides interacting with all its neighbors. We obtain the general equations for the couplings to fulfill the normalization condition and match the arguments of the Fourier matrix. From these equations, we provide an analytical solution for $N=6$ and determine where our model is subdued by experimental feasibility.

\begin{figure}[t!]
    \centering
    \includegraphics[width=\linewidth]{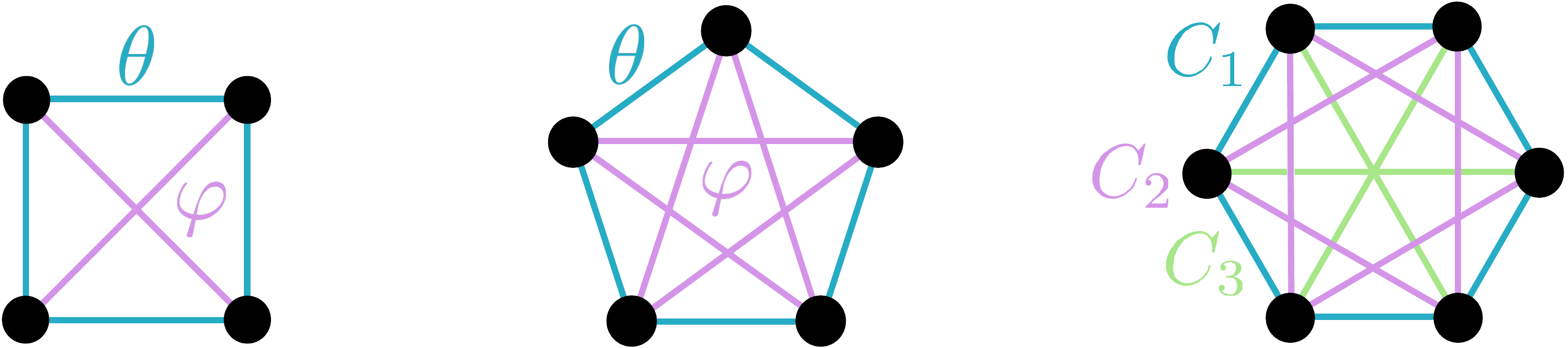}
    \caption{Regular polygonal geometries representing the cross section of multiarm interferometers. The colors indicate the different distances in the polygon. $\theta$, $\varphi$, and $C_{1,2,3}$ are the couplings.}
    \label{fig:polygon23456}
\end{figure}

From the perspective of graph theory, the coupling matrices we have studied so far correspond to a complete graph, that is, a graph where every vertex is connected to every other vertex. The vertices represent the waveguides and the edges the couplings. Ahmadi \textit{et al}. \cite{Ahmadi2003} demonstrated that instantaneous uniform mixing, that is, having $|U_{m,n}^{(N)}|=1/\sqrt{N}$ \cite{Chan2020}, is possible only for $N=2,\ 3,\ 4$ in complete unweighted graphs. This result implies that we have to relinquish the idea of finding the DFT through complete-graph-like coupling matrices with a unique coupling coefficient. However, in Eq.~\eqref{eq:DFT4_phases} we found that the $4$-dimensional QFTI was accessible with two different couplings, which strongly suggests examining the $5$-dimensional case in the same manner. By doing this, we achieved the $5$-dimensional QFTI from the coupling matrix
\begin{align}\label{eq:coupling_5x5}
    \mathcal{C} =
    \begin{pmatrix}
        0 & \theta & \varphi & \varphi &\theta \\
        \theta & 0 & \theta & \varphi & \varphi\\
        \varphi & \theta & 0 & \theta & \varphi \\
        \varphi & \varphi & \theta & 0 & \theta\\
        \theta & \varphi & \varphi & \theta & 0
    \end{pmatrix},
\end{align}
which represents the interactions in the pentagon arrangement (see Fig.~\ref{fig:polygon23456}) and is equivalent to a complete weighted graph. The complex exponential of $\mathcal{C}$ in Eq.~\eqref{eq:coupling_5x5} provides the squares moduli
\begin{subequations}\label{eq:U_mn(5x5)}
    \begin{align}
    \nonumber\left| U_{1,1}^{(5)}\right|^2
    &=\frac{1}{25}\left[9+8\cos{\left(\sqrt{5}(\theta -\varphi)/2\right)}\cos{\left(5(\theta +\varphi)/2\right)}\right.\\
    &\phantom{=}\left. +8\cos{\left(\sqrt{5}(\theta -\varphi)\right)}\right],\\
    \nonumber \left| U_{1,2}^{(5)}\right|^2 &=\frac{2}{25}\left[\vphantom{\sqrt{5}} 2+\sqrt{5}\sin{\left(5(\theta+\varphi)/2\right)}\sin{\left(\sqrt{5}(\theta -\varphi)/2 \right)}\right.\\
    \nonumber &\phantom{=}-\cos{\left(5(\theta+\varphi)/2\right)}\cos{\left(\sqrt{5}(\theta -\varphi)/2 \right)}\\
    &\left. \phantom{=} -\cos{\left(\sqrt{5}(\theta -\varphi)\right )}\right] ,\\
    \nonumber \left| U_{1,3}^{(5)}\right|^2  &=\frac{2}{25}\left[ \vphantom{\sin{\left(\sqrt{5}\right)}}2-\sqrt{5}\sin{\left(5(\theta+\varphi)/2\right)}\sin{\left(\sqrt{5}(\theta -\varphi)/2 \right)}\right.\\
    \nonumber &\phantom{=}-\cos{\left(5(\theta+\varphi)/2\right)}\cos{\left(\sqrt{5}(\theta -\varphi)/2 \right)}\\
    &\left.\phantom{=}-\cos{\left(\sqrt{5}(\theta-\varphi)/2\right)}\right],
\end{align}
\end{subequations}
where $U_{14}^{(5)}=U^{(5)}_{13}$, $U_{15}^{(5)}=U_{12}^{(5)}$ and the remaining rows of $U^{(5)}$ are cyclically shifted to the right from the preceding. By imposing the normalization condition on Eqs.~\ref{eq:U_mn(5x5)}, that is, $|U_{1k}^{(5)}|^2=1/5$, the system is reduced to
\begin{subequations}\label{eq:pentagon}
    \begin{align}
        0 &=\sin{\left(\frac{\sqrt{5}}{2}(\theta -\varphi)\right)}\sin{\left(\frac{5}{2}(\theta+\varphi)\right)}, \\
        0 &= \cos^2{\left(\frac{\sqrt{5}}{2}(\theta-\varphi)\right)}-\frac{1}{2}\cos{\left(\frac{\sqrt{5}}{2}(\theta-\varphi)\right)}-\frac{1}{4}.
    \end{align}
\end{subequations}

The set of all solutions of Eq.~\eqref{eq:pentagon} is given by 
\begin{subequations}\label{eq:5x5_solutions}
    \begin{align}
        \theta &= \frac{\pi}{25}\left(5k_1+n\sqrt{5}+2k_2\right),\\
        \varphi &= \frac{\pi}{25}\left(5k_1-n\sqrt{5}-2k_2\right),
    \end{align}
\end{subequations}
where $n=1,\ 3\ (2,\ 4)$ if $k_1$ is odd (even) and $k_{1,2}\in \mathbb{Z}$.

It is important not to overlook the choice of integers $k_{1,2}$ and $n$, as this choice determines both the interaction length and the coupling strength. A non-intuitive consequence also arises: the presence of permutation matrices. If we take $k_1=2$, $k_2=0$, $n=2$ we need the permutations of columns $P=[\hat e_1, \hat e_3,\hat e_5, \hat e_2, \hat e_4]$, where $\hat e_j$ is the canonical base, however, taking $k_1=1$, $k_2=0$ and $n=1$, no permutation matrices are required. In the last case, the respective phase-shifter matrices are given by
\begin{subequations}
    \begin{align}
        \Phi^{\rm in} &={\rm diag}\left(e^{-i\pi/5},\ e^{3i\pi/5},\ e^{i \pi},\ e^{i\pi},\ e^{3i\pi /5} \right),\\
        \Phi^{\rm out} &= {\rm diag}\left(1,\ e^{4i\pi /5},\ e^{-4i\pi /5},\ e^{-4i\pi /5},\ e^{4i\pi /5}\right).
    \end{align}
\end{subequations}

The previous result for the $5$-dimensional QFTI motivated us to examine a general polygonal model. Since the distances between the vertices of a regular polygon are cyclic, the general coupling matrix is circulant, that is, all rows are composed of the same elements, and each row is rotated one element to the right relative to the preceding row. Thereby, the coupling matrix becomes
\begin{align}\label{eq:Coupling_polygon}
    \mathcal{C} &=
    \begin{pmatrix}
    0 & C_1 & C_2 & \cdots & C_{\mathsmaller{N}-2} & C_{\mathsmaller{N}-1} \\
    C_{\mathsmaller{N}-1} & 0 & C_1 & \cdots & C_{\mathsmaller{N}-3} & C_{\mathsmaller{N}-2} \\
    \vdots & \vdots & \vdots & \ddots & \vdots & \vdots \\
    C_{2} & C_{3} & C_{4} & \cdots & 0 & C_1\\
     C_1 & C_2 & C_3& \cdots & C_{\mathsmaller{N}-1} & 0
    \end{pmatrix}.
\end{align}
Circulant matrices are diagonalized through the discrete Fourier matrix $F$, allowing us to determine $U^{(N)}$ as
\begin{align}
    e^{-i\mathcal{C}z} &=F^{\dagger}e^{-i\Lambda z} F,
\end{align}
where $\Lambda$ is the diagonal matrix that contains the eigenvalues $\lambda_k$ of $\mathcal{C}$. Consequently, the coefficients of $U^{(N)}$ are expressed as
\begin{align}\label{eq:U_mn}
    U_{m,n}^{(N)} &= \frac{1}{N} \sum\limits_{k=0}^{N-1}\exp{\left(2\pi i k(n-m)/N-i\lambda_k\right)}.
\end{align}

\begin{figure}[ht!]
    \centering
    \includegraphics[width=\linewidth]{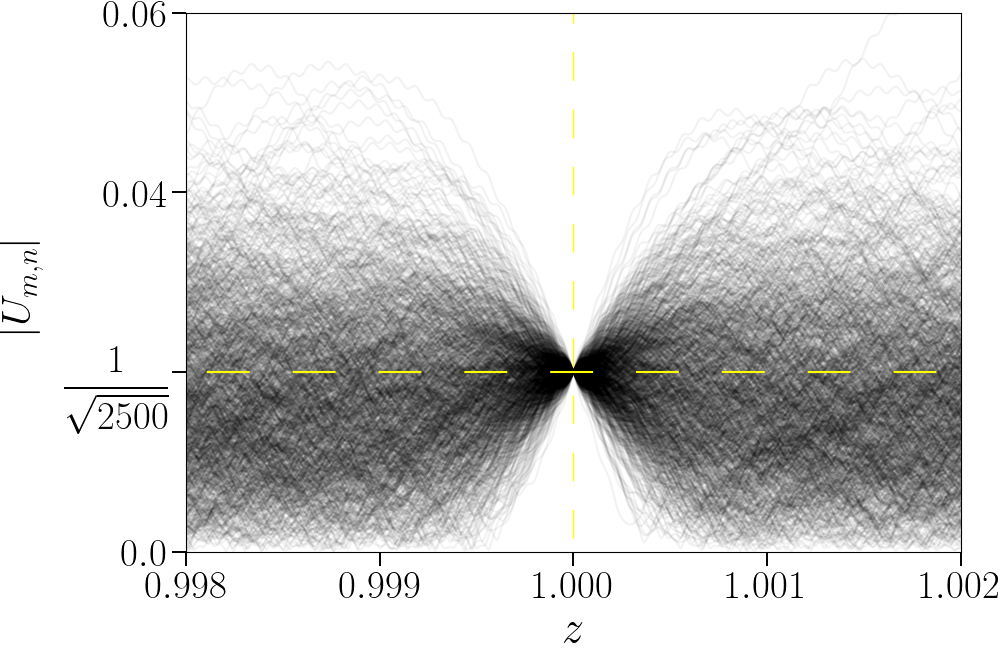}
    \caption{Moduli of the entries for $U_{2500}$ as function of the interaction length. The yellow dashed lines indicates the point where the balancing condition is met.}
    \label{fig:Hadamard_2500}
\end{figure}

In a regular polygon of $S$ sides, we have $\lfloor S/2 \rfloor$ different distances, so the evanescent coupling strength felt by the waveguides has symmetry $C_l=C_{N-l}$ for $l=1,\ldots,N-1$. Thus, the eigenvalues of $\mathcal{C}$~\eqref{eq:Coupling_polygon} are given by
\begin{subequations}
    \begin{align}
    \label{eq:lambda_odd}\lambda_k &=C_0+2\sum\limits_{l=1}^{(N-1)/2} C_l \cos{\left( 2\pi k l /N \right)},\;\\
    \label{eq:lambda_even}\lambda_k &=C_0+(-1)^kC_{N/2}+2\sum\limits_{l=1}^{N/2-1} C_l\cos{(2\pi kl/N)}.
    \end{align}
\end{subequations}
Eq.~\eqref{eq:lambda_odd} and Eq.~\eqref{eq:lambda_even} correspond to the eigenvalues when $N$ is odd or even, respectively. They have symmetry $\lambda_k=\lambda_{N-k}$ for $1\leq k \leq \lceil N/2\rceil-1$.

The normalization condition over $U_{mn}^{(N)}$~\eqref{eq:U_mn} is captured by the following equation (see Appendix~\ref{Appendix:norm_angles} for details)
\begin{align}\label{eq:norm_equation}
    0=\sum\limits_{j=1}^{N-1}\sum\limits_{k=j}^{N-1}\cos{\left(\frac{2\pi j(m-n)}{N}+\lambda_k-\lambda_{k-j}\right)}.
\end{align}
A solution to Eq.~\eqref{eq:norm_equation} does not guarantee that we have found the DFT, but would give rise to a complex Hadamard matrix (CHM)~\cite{Dita2004}. To ensure that $U^{(N)}$~\eqref{eq:U_mn} is the $N$-dimensional QFTI, the core of the dephased form of $U^{(N)}$, that is, the submatrix $U_{jk}^{(N)}$, $j,k=1,\ldots, N-1$ (core) when $U_{0,j}^{(N)}=U_{j,0}^{(N)}=1$ (dephased), must coincide with Eq.~\eqref{eq:dft}. This constrains the solutions of Eq.~\eqref{eq:norm_equation} to also fulfill the equation (see Appendix~\ref{Appendix:norm_angles} for details)
\begin{align}\label{eq:angle_equation}
    \nonumber 0 &= \sum\limits_{j,k,l,p=0}^{N-1} \sin{\left((k-l)\frac{2\pi n}{N} +(j-k)\frac{2\pi m}{N}+\lambda_{jklp}\right)}\\
    &-T_{nm}\cos{\left((k-l)\frac{2\pi n}{N} +(j-k)\frac{2\pi m}{N}+\lambda_{jklp}\right)},
\end{align}
where $T_{mn}=\tan{\left(\frac{2\pi nm}{N}\right)}$ and $\lambda_{jklp}=\lambda_j-\lambda_k+\lambda_l+\lambda_p$. The dephased version of $U^{(N)}$, which is denoted by $\mathcal{D}^{(N)}$, is then given by
\begin{subequations}\label{eq:U_dephased}
    \begin{align}
        \mathcal{D}^{(N)} &= \Phi^{\rm out}U^{(N)}\Phi^{\rm in},\\
        \Phi^{\rm in} &= {\rm diag}\left( u_{0,0}^{(N)\ast},\ldots,u_{0,n}^{(N)\ast}, \ldots, u_{0,N-1}^{(N)\ast}\right),\\
        \Phi^{\rm out} &={\rm diag}\left(1, u_{1,0}^{(N)\ast},\ldots,u_{m,0}^{(N)\ast},\ldots,u_{N-1,0}^{(N)\ast }\right),
    \end{align}
\end{subequations}
where $u_{mn}^{(N)}=\sqrt{N} U_{mn}^{(N)}$.

\begin{figure}[ht!]
    \centering
    \includegraphics[width=\linewidth]{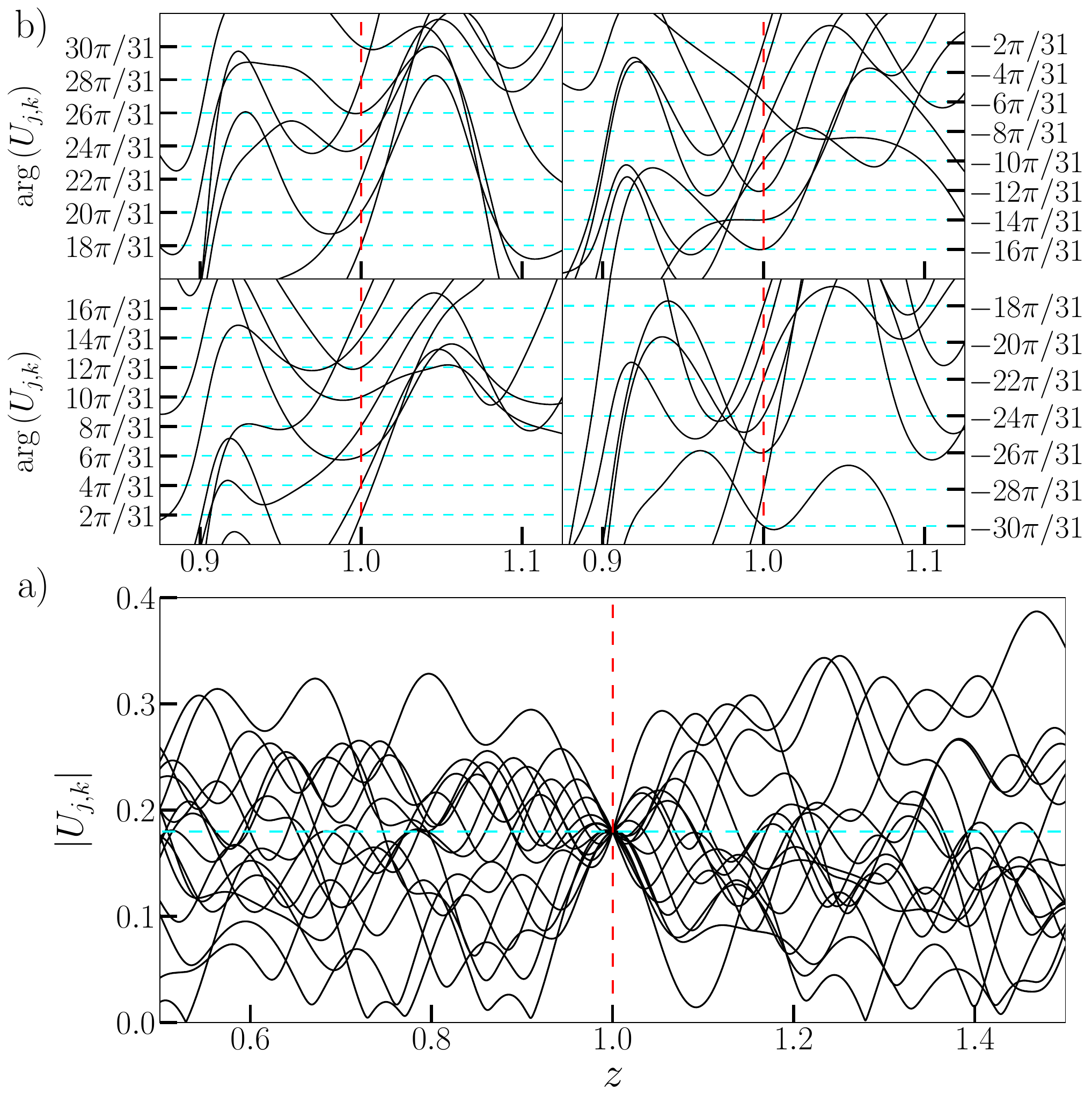}
    \caption{a) Moduli of $U$ for $N=31$ solved from Eq.~\eqref{eq:norm_equation} and checked in Eq.~\eqref{eq:angle_equation}. b) Arguments of $U$ as function of the interaction length $z$. The intersections of the blue dashed lines with the red vertical dashed line shows the argument matching of our solution with the DFT.}
    \label{fig:DFT_31}
\end{figure}

It is important to note that $\exp{(-i\mathcal{C}z})$ inherits the same circulant structure $\mathcal{C}$ has~\eqref{eq:Coupling_polygon}, and therefore we only need to solve Eq.~\eqref{eq:norm_equation} for $m=0$ and $n=0,\ldots, \lfloor (N-1)/2\rfloor$. Due to the circulant structure of $\mathcal{C}$, the unitarity of $U=e^{-i\mathcal{C}}$ imposes only one constraint, resulting in a non-linear square system of $\lfloor N/2 \rfloor$ independent equations. However, Eq.~\eqref{eq:angle_equation} introduces other $\lfloor N/2\rfloor+1$ equations for couplings, making the system of equations to find the DFT~\eqref{eq:dft} overdetermined. Nevertheless, we solved Eq.~\eqref{eq:norm_equation} (see the reduced version in Eqs.~\eqref{eq_appe:Norm_even} and \eqref{eq_appe:Norm_odd} for $N$ even and odd, respectively) numerically with machine precision and found solutions up to $N=31$ for the DFT (see Fig.~\ref{fig:DFT_31}), and up to $N=2500$ of being a CHM (see Fig.~\ref{fig:Hadamard_2500}). Whether the polygonal model can always yield solutions to obtain $F_N$ from Eq.~\eqref{eq:Coupling_polygon} or not is beyond our theoretical efforts.

The main drawback that the polygonal model has in PICs is the exponential decay of the evanescent coupling strength with the distance between non-adjacent waveguides as $N$ increases~\cite{Szameit2007}. Consequently, the question to answer is to what extent this model is suitable for experimental realization in 3D waveguides. 
The pentagon only requires second order couplings (next nearest neighbor) at $1.62 a$, where $a$ is the distance from center to center, and this is experimentally feasible~\cite{Dreisow2008,Szameit2007}. 
For the hexagon, its geometry requires a third order coupling at twice the distance of the first neighbors (next-to-next nearest neighbor). The proportion of first and third order coupling, $C_1/C_3$, following an exponential decay of the coupling strength with distance, $C\propto e^{-\gamma x}$, requires $\gamma = \ln(C_1/C_3)/a$. From Eq.~\eqref{eq:norm_equation} we found $C_1=\pi/3$, $C_2=\pi/6$ and $C_3=\pi/12$ (see Appendix~\ref{Appendix:DFT6}), therefore $\gamma=\ln(4)/a$. This implies that we need $a$ as short as possible, so this case is on the frontier of experimental capabilities. Taking waveguides with diameter $7\ \mu{\rm m}$~\cite{Xuhu_Han2024} and a gap width of $7\ \mu{\rm m}$~\cite{Yuying_Wang2025}, we have $a=14\ \mu {\rm m}$, thereupon $\gamma=0.1\ \mu{\rm m}^{-1}$, which is experimentally achievable~\cite{Lijing_Zhong2025}. We can conclude that the polygonal model is bound to $N\leq6$, since for the heptagon, the longest diagonal would have a length of $\approx 2.25 a$.

For the solution we have presented for the hexagon, the phase-shifter matrices required to obtain $F_6$ are given by
\begin{subequations}
    \begin{align}
    \Phi^{\rm in} &= {\rm diag}\left(e^{-i\frac{5\pi}{18}}, e^{i\frac{5\pi}{9}}, e^{-i\frac{17\pi}{18}}, e^{-i\frac{7\pi}{9}}, e^{-i\frac{17\pi}{18}}, e^{i\frac{5\pi}{9}}\right),\\
    \Phi^{\rm out} &= {\rm diag}\left(1,e^{i\frac{5\pi}{6}},e^{-i\frac{2\pi}{3}},e^{-i\frac{\pi}{2}},e^{-i\frac{2\pi}{3}},e^{i\frac{5\pi}{6}}\right).
\end{align}

\end{subequations}

\section{\label{sec:new_DFT} Graph-based QFTI}

Multiarm interferometers can be seen as a graph, where the vertices correspond to the waveguides, and the edges represent the couplings. In our case, all possible interferometers are mapped to connected graphs on $N$ unlabeled vertices, \textit{i.e.}, graphs where every vertex has at least one edge (connected) and the labels can be arbitrary (unlabeled).

In this section, to address the inexorable limitations of the polygonal model regarding the decaying coupling strength with the distance between waveguides, we explore the coupling configurations according to the connected unlabeled graphs in the case of non-complete graphs, \textit{i.e.}, graphs with at most $N(N-1)/2-1$ edges. We show that different propagation constants $\beta_l \neq 0 $ in the Hamiltonian are necessary to obtain the $N$-dimensional QFTI from these new cases. We establish a numerical procedure to deal with the free phases in the evolution operator in order to acquire the specific solutions that lead to the Fourier matrix~\eqref{eq:dft}.

\subsection{\label{sec:numerical_CHM} Numerical approach and CHMs}

We start this subsection showing how the inclusion of propagation constants $\beta_l\neq 0$ allows one to obtain the $N$-dimensional QFTI from a waveguide array where with $\beta_l=0$ is not possible. Then we present the numerical strategy to deal with the balancing condition. From this numerical procedure, CHMs will naturally emerge, and we deepen in this subject on the relevant properties concerning the DFT~\eqref{eq:dft}.

For three waveguides, the two possible geometrical configurations are the tritter and the nearest-neighbor line. The tritter leads to a QFTI~\cite{Spagnolo2013}, the line does not, since it is not possible to balance the moduli in its transformation matrix even when the couplings are different (see Appendix C in Ref.~\cite{Rojas2019}). Notwithstanding, if we take the propagation constant $\beta$ into account, the $3$-dimensional QFTI is within reach for the nearest-neighbor line when we assign $\beta\neq0$ to the central waveguide. The demonstration of moduli balancing to $1/\sqrt{3}$ and phase matching with the roots of unity for this transformation can be found in the Appendix~\ref{Appendix:new_Fourier3}. Unlike the tritter, a permutation between the second and third column is now necessary. Although this new configuration for the $3$-dimensional QFTI would require more engineering in the optical circuit, incorporation of the propagation constant into the bosonic evolution reveals that $\beta_l$ is another relevant variable in the system in the course of finding the $N$-dimensional QFTI through a single stage of evolution.

To find the couplings that lead to the $N$-dimensional QFTI for $N\geq 4$ waveguides, we use numerical calculations, since as the number of couplings increases and different propagation constants are taken into account, analytical solutions become intractable. To address the problem, we construct a real multivariate vector-valued function $\vec f$ defined as the vectorization
\begin{align}\label{eq:f_numerical}
    \vec f(\{\beta_l,C_{jk}\}) &= {\rm vec}_{\{\rm sym \}}\left(  e^{-i\mathcal{C}}\odot e^{i\mathcal{C}}-\frac{1}{N}\mathbb{J}\right),
\end{align}
where $\odot$ indicates the Hadamard or entrywise product and $\mathbb{J}$ is the all-ones matrix. To avoid repeated equations in vectorization $\vec f$, we only conserve 
unique elements $|U_{jk}|^2$. This is denoted as $\{{\rm sym\}}$ and is determined by the symmetries of $C_{jk}$ and $\beta_l$.

In order to find the couplings, we sought the roots of $\vec f(\{\beta_l, C_{jk}\})$, \textit{i.e.}
\begin{align}\label{eq:f_numerical=0}
    \vec f &= \left(\left|U_{11}\right|^2-1/N,\ldots,|U_{jk}|^2-1/N,\ldots\right)= \vec 0,
\end{align}
which is equivalent to the normalization condition $|U_{jk}|=1/\sqrt{N}$, but numerically more stable. To solve Eq.~\eqref{eq:f_numerical=0}, we implemented the Levenberg–Marquardt algorithm through the \textit{root} function of the SciPy library (v. 1.15.1)~\cite{Virtanen2020}. To guaranty an accurate solution, after solving $\vec f =\vec 0$~\eqref{eq:f_numerical=0} we verified the convergence of the Euclidean norm of $\vec f$ to zero within machine precision.

The evolution of annihilation and creation operators, $U$~\eqref{eq:U=exp_C}, is unitary by construction, and a solution to Eq.~\eqref{eq:f_numerical=0} ensures balanced entries in $U$. These are the two conditions that define the CHMs, from which the DFT~\eqref{eq:dft} is just a particular case, where the matrix elements are the roots of unity $\omega_k = e^{-2i\pi k/N}$. Nothing guaranties that a solution of Eq.~\eqref{eq:f_numerical=0} will converge to a solution with these phases; moreover, there can be continuous parameters in $U_{jk}$, like the phase $\phi$ in Eq.~\eqref{eq:Fourier_family_4x4}. Therefore, we also have to carefully examine the phases in $U_{jk}$.

To numerically analyze the phases within $U$ two concepts from the study of CHMs are relevant to us: the defect and the affine families. The defect of a CHM $H$, denoted as $d(H)$, is an upper bound on how many free phases (like $\phi$ in Eq.~\eqref{eq:Fourier_family_4x4}) there can be in $H$. When $d(U)\neq d(F_N)$, $U$ is not $F_N$ certainly, however, if $d(U)=d(F_N)$, we cannot say. When this happens, the affine families come into play as we need to analyze how the phases are organized in the matrix. An affine family refers to the entire structure of the matrix in which the number and location of the free phases are characterized. Affine stands for phases with linear dependence. In what follows, we deepen the discussion with more rigorous definitions for these concepts and their role in the numerical routine to find $F_N$. 

\textbullet\ The defect is an algebraic invariant under the equivalence of CHMs, that is, when there exist permutations $P_{1,2}$ and diagonal matrices $D_{1,2}$ that hold the equality
\begin{align}
    H_1 &= P_{1} D_{2} H_2 D_{2}P_{2},
\end{align}
where $H_{1,2}$ are the CHMs.

The defect corresponds to the dimension of the solution space of the system of equations
\begin{align}\label{eq:defect}
    0 &= \sum\limits_{l=0}^{N-1}H_{kl}H^{\ast}_{jl}\left(\alpha_{kl}-\alpha_{jl}\right),
\end{align}
where $\alpha_{jk}$ are the arguments of complex exponentials perturbing the entries of $H$, and we will refer to them as free phases. The dephased condition imposes constraints $0=\alpha_{0j}$ ($j=1,\ldots,N-1$) and $0=\alpha_{k0}$ ($k=0,\ldots,N-1$) on the factors $e^{i\alpha_{jk}}$, reducing the rank of Eq.~\eqref{eq:defect} in $2N-1$. Equation~\eqref{eq:defect} comes from the differentiation of the equations behind the orthogonality of the rows (or columns) in a CHM, namely
\begin{align}
    0 &= \sum\limits_{l=0}^{N-1} H_{kl}H^{\ast}_{jl}e^{i(\alpha_{kl}-\alpha_{jl})}.
\end{align}

More details about this algebraic invariant can be consulted in Ref.~\cite{Tadej2006,Bengtsson2007,Barros2013}. In the case of the Fourier matrix, the defect is known for any $N$ (see Theorem 5.2 in Ref.~\cite{Tadej2008}) and is given by
\begin{align}\label{eq:fourier_defect}
    d(F_N) &= 
    \begin{cases}
        1-N+2\sum\limits_{l=1}^{\frac{N-1}{2}}{\rm gcd}(N,l) \quad\;\;\:  N\  {\rm odd}, \\
        1-\frac{N}{2}+2\sum\limits_{l=1}^{\frac{N}{2}-1} {\rm gcd}(N,l) \quad\:\:\: N\ {\rm even},
    \end{cases}
\end{align}
where ${\rm gcd}(N,l)$ is the greatest common divisor. In the numerical routine, after solving Eq.~\eqref{eq:f_numerical=0}, we have to calculate the defect of the corresponding dephased version of $U$ ($\Phi^{\rm out} U\Phi^{\rm in}$) or subtract $2N-1$ from $d(U)$ and compare it with $d(F_N)$~\eqref{eq:fourier_defect}. If $d(\Phi^{\rm out} U\Phi^{\rm in})$ is different from the value provided by Eq.~\eqref{eq:fourier_defect}, then $U$ is not the DFT. However, if they are equal, $U$ is not necessarily the DFT, and we must examine the structure of the phases within $U$ according to the corresponding affine family of $F_N$.

\textbullet\ An affine family of CHMs are the matrices stemming from the continuous factors $H_{jk}e^{i\alpha_{jk}}$ that do not alter the Hadamard condition of the matrix and where the phases $\alpha_{jk}$ only have linear dependence with each other, that is, $\alpha_{jk} = \sum_{\{m,n\} } \alpha_{mn}$. Consequently, their structure is as follows
\begingroup
\setlength\arraycolsep{1.9pt}
\begin{align}
    H \!&=\!\frac{1}{\sqrt{N}}\!\!
    \begin{pmatrix}
        1 & 1  & \cdots &  1 \\
        1 & H_{1,1}e^{i\alpha_{1,1}} & \cdots & H_{2,N-1}e^{i\alpha_{1,N-1}}\\
        \vdots & \vdots  & \ddots & \vdots\\
        1 & H_{N-1,2}e^{i\alpha_{N-1,2}} & \cdots & H_{N-1,N-1}e^{i\alpha_{N-1,N-1}}
    \end{pmatrix}\!.
\end{align}
\endgroup

If the relation among the phases $\alpha_{jk}$ is not linear, then it is called a non-affine family. A CHM can contain both affine and non-affine arguments~\cite{Karlsson2009}. When a dephased CHM, say $H$, is fully described by a set of affine phases, and this set cannot be larger, the affine family stemming from $H$ is called a maximal affine Hadamard family. For example, for the Fourier matrix in dimension $N=4$ its affine family only has the phase $\phi$, as can be appreciated in Eq.~\eqref{eq:Fourier_family_4x4}. However, in dimension $N=6$ the defect is $4$, and the maximal affine Hadamard family only has two phases~\cite{Tadej2006}. In Sec.~\ref{sec:qfti_N=6} we will see that the other two parameters correspond to non-affine phases.

The structure of the affine families plays an important role in the search for the DFT in the numerical routine because the value of the defect does not verify the inequivalence between the CHMs when $d(H_1)=d(H_2)$. To distinguish whether the transformation $U$ we obtained from Eq.~\eqref{eq:f_numerical=0} is the DFT, there are elements with fixed phases in the corresponding affine family of $F_N$ (rows and columns whose indices share one divisor with $N$) and we look for them in $U$, checking the difference $\lvert \arg(U_{jk})-2\pi jk/N \mod(2\pi)\rvert\approx 0 $ within machine precision.  For prime $N$ the defect of $F_N$ is zero, and therefore the arguments of $U$ can be compared with $F_N$ without ambiguities. We must bear in mind that permutations may be present, like for the pentagon in the polygonal model, so the indices in the dephased version of $U$ do not necessarily coincide with $F_N$.

All connected unlabeled graphs that we use throughout this work were generated using the program \textit{geng} from the \textit{Nauty} \& \textit{Traces} software package (version 2.9)~\cite{McKay1998, MCKAY2014,McKay2026}. For $N=4,\ 5,\ 6,\ 7,\ 8,\ 9,\ 10$ the number of graphs are $6$, $21$, $112$, $853$, $11117$, $261080$, $11716571$, respectively. Given the large number of graphs to study, in each numerical search we considered the solution to be a success when the transformation $U$ from the solution of Eq.~\eqref{eq:f_numerical=0} had the same defect as the Fourier matrix~\eqref{eq:fourier_defect} and the structure of the corresponding affine family coincided. To solve Eq.~\eqref{eq:f_numerical=0} numerically, we used random guesses for $C_{jk}$ and $\beta_l$ with a uniform distribution, this being the bottleneck in the routine.

\begin{figure}[ht!]
    \centering
    \includegraphics[width=0.9\linewidth]{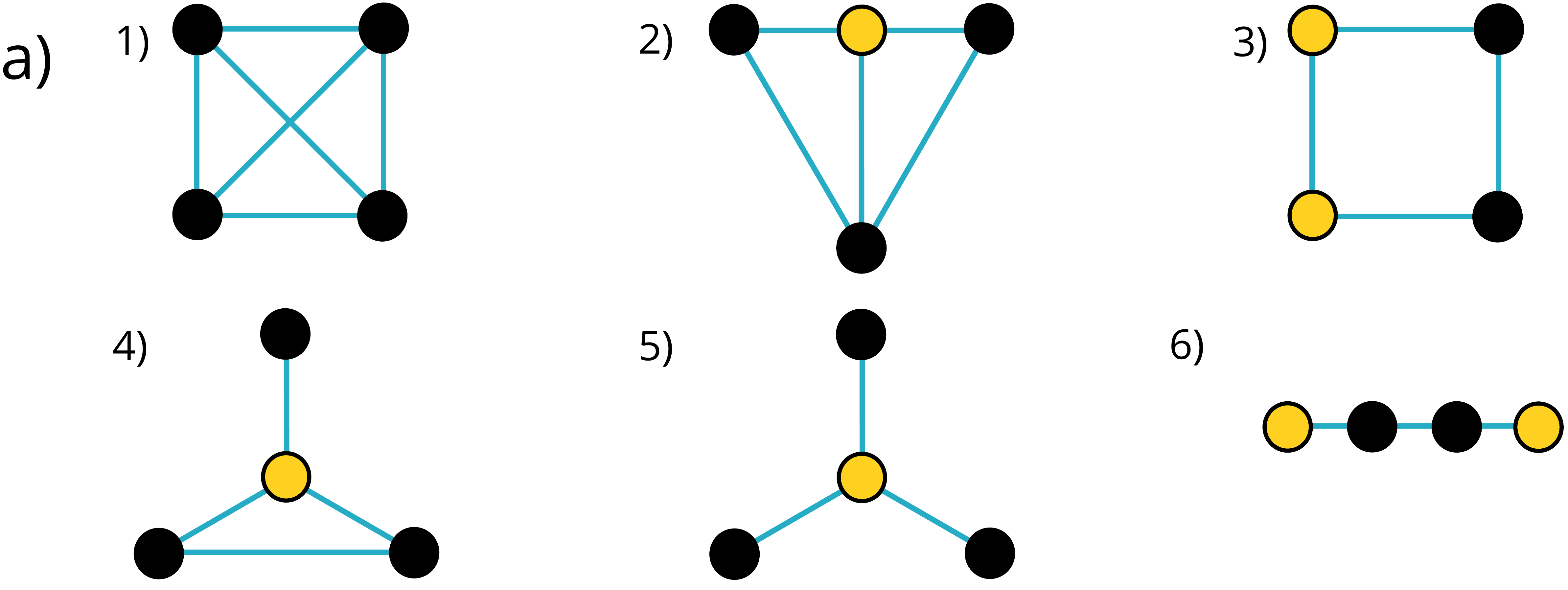}
    \includegraphics[width=\linewidth]{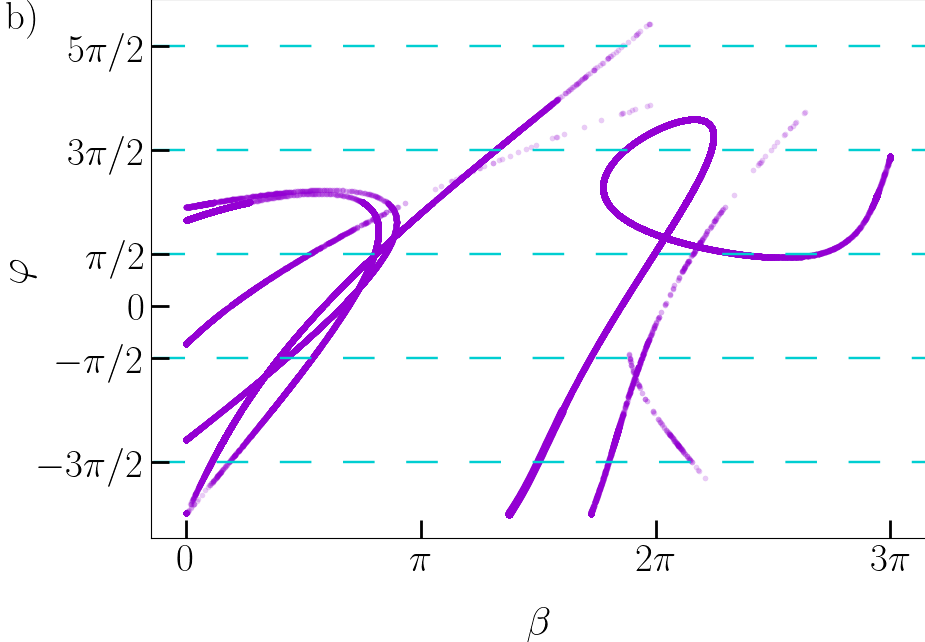}
    \caption{a) Cross section of the six arrangements of four waveguides. The black circles represent the waveguides and the blue lines amongst them, the couplings. Yellow circles indicates waveguides with propagation constants $\beta_l\neq 0$. b)~Free phase $\varphi$ in Eq.~\eqref{eq:Fourier_family_4x4} as function of $\beta$ for arrays $4)$. The intersection of $\varphi(\beta)$ with $\pm\pi/2$, $\pm3\pi/2$ and $5\pi/2$ confirm the possibility of achieving the $4$-dimensional QFTI.}
    \label{fig:qfti4}
\end{figure}

\subsection{\label{sec:prime_and_affine} Prime dimension and maximal affine families}

In this subsection, we present the graphs that lead to a QFTI for prime $N$ and composite $N$ when the family of the Fourier matrix is maximally affine.

The six geometrical configurations of four waveguides are shown in Fig.~\ref{fig:qfti4} a). First, we considered $\beta_l=0$, and only the nearest-neighbor line (case $6)$ in Fig.~\ref{fig:qfti4} a) ) did not satisfy normalization to $1/2$. The cases $2)$, $3)$, $4)$ and $5)$ converged to Eq.~\eqref{eq:Fourier_family_4x4} for specific values of the free phase $\phi$, but not for $\phi=\pm\pi/2$, which is the value that determines $F_4$ in Eq.~\eqref{eq:Fourier_family_4x4}. To complete the analysis for $N=4$, we included $\beta_l\neq0$ as indicated in Fig.~\ref{fig:qfti4} a). In case $6)$, the solutions for $C_{jk}$ and $\beta_l$ satisfying the balancing condition directly determined the $4$-dimensional QFTI. Both propagation constants and the three couplings were not equal.

For cases $2)$, $3)$, $4)$ and $5)$ the search for $F_4$ was not direct, since Eq.~\eqref{eq:f_numerical=0} only captures the square moduli in the entries of $\exp{(-i\mathcal{C}})$ and, as can be appreciated in Eq.~\eqref{eq:Fourier_family_4x4}, the free phase $\phi$ is a continuous variable that does not play a role in normalization. To reveal the behavior of $\phi$ as a function of $\beta_l$, we had to solve Eq.~\eqref{eq:f_numerical=0}  $\approx 10^{6}$ of times. The numerical success rate was $\sim 66\%,\ 50\%,\ 16\%,\ 39\%$ for cases $2)$, $3)$, $4)$ and $5)$ in Fig.~\ref{fig:qfti4} b), respectively. In $3)$ the propagation constant is the same for both waveguides and in $5)$ they are not. In Figure.~\ref{fig:qfti4} b) we can observe the non-trivial behavior of $\phi$ in terms of $\beta_l$, and the value $\phi=\pm\pi/2$ confirmed the possibility of obtaining $F_4$ from configurations $2)$, $3)$, $4)$ and $5)$. It is worth mentioning that $\phi$ also depends on the couplings $C_{jk}$, but the propagation constant is the variable that allows $U$ to become the DFT. For this reason, we studied $\phi$ as a function of $\beta_l$.

\begin{figure}[ht!]
    \centering
    \includegraphics[width=1\linewidth]{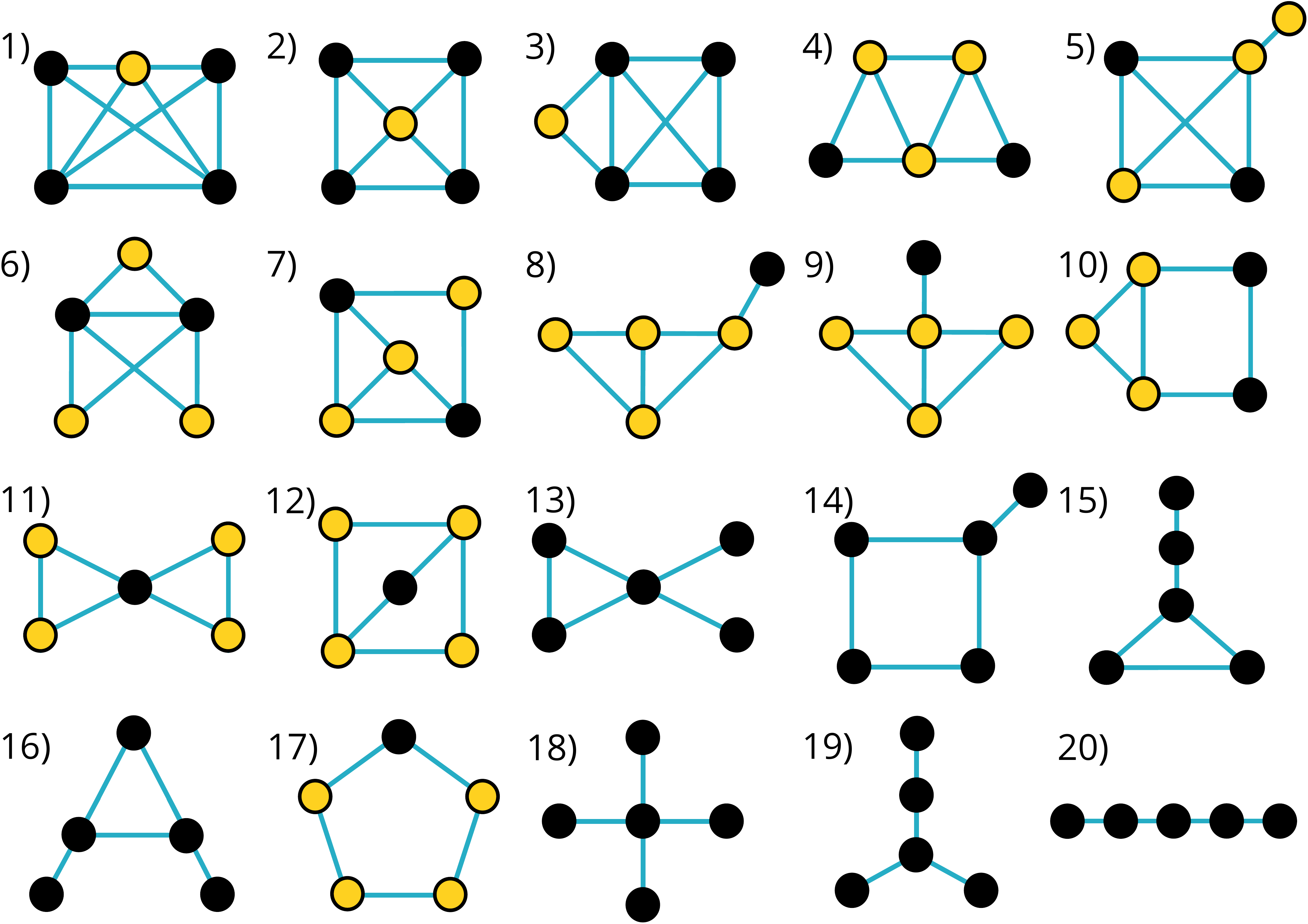}
    \caption{Twenty possible coupling configurations of five waveguides (cross section). Black circles: waveguides; Yellow circles: $\beta\neq0$; blue lines: couplings.}
    \label{fig:M5_all}
\end{figure}

There are a total of $21$ connected unlabeled graphs of five vertices, in other words, $21$ possible geometrical configurations for the multiarm interferometers. In Sec.~\ref{sec:polygonal_model} we demonstrated that a pentagonal geometrical configuration of waveguides is capable of implementing $F_5$. In Fig.~\ref{fig:M5_all} we can see the remaining $20$ waveguide dispositions. In this dimension, all CHMs are equivalent to $F_5$~\cite{Tadej2006}, hence, we do not need to analyze the arguments in $U$ once it is dephased by the proper phase-shifter matrices. Disregarding the propagation constant, none of these $20$ candidates guided us to the DFT. Nonetheless, with $\beta_l\neq 0$ included in the coupling matrix, the arrangements from $1)$ to $12)$ and $17)$ shown in Fig.~\ref{fig:M5_all} converged to $F_5$.

\begin{figure}
    \centering
    \includegraphics[width=\linewidth]{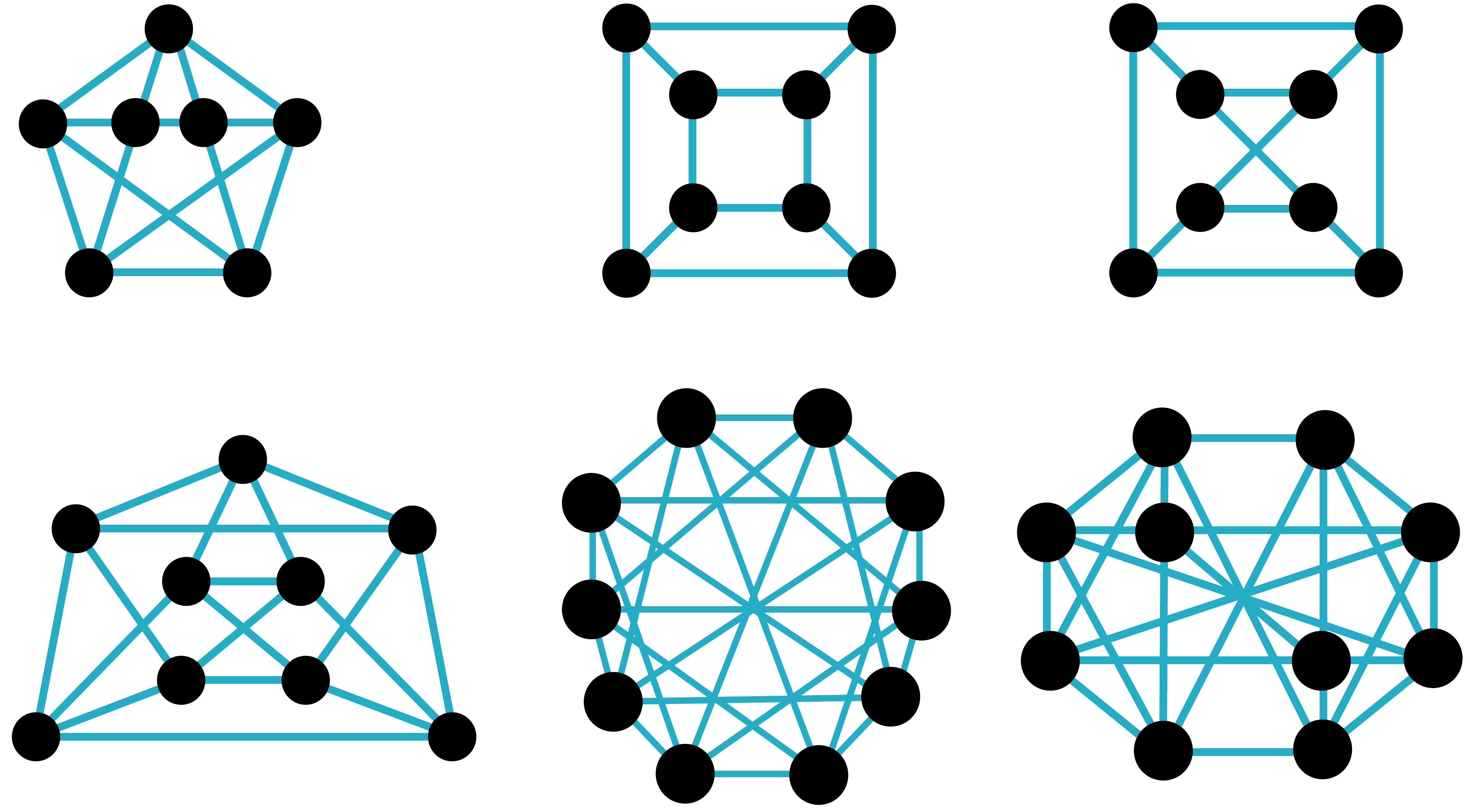}
    \caption{Coupling configurations with an associated regular graph. Black circles: waveguides; blue lines: couplings. From left to right and top to bottom: $4$-, $3$-, $4$-, and $5$- regular graphs, for $N=7,\ 8,\ 9,\ 10$, respectively.}
    \label{fig:regular}
\end{figure}

As we stated earlier, when $N$ is prime, we have $d(F_N)=0$. This means that there are no free phases involved, the arguments in $F_N$ are fixed, and the comparison between $U$ and $F_N$ is straightforward, aside from permutations. Owing to the existence of other CHMs not equivalent to $F_7$~\cite{Tadej2006}, in this case, albeit simple, we cannot skip this step of the analysis. From the $853$ candidates to the $7$-dimensional QFTI, we found $80$ solutions (heptagon not counted), and they correspond to the graphs with a number of edges $\lvert E\rvert$ in the range $15\leq \lvert E\rvert\leq 20$ and one graph with $14$ edges. The latter is a $4$-regular graph (see Fig.\ref{fig:regular}). The $k$-regular graphs are a special kind of graph where every vertex has the same number of edges.

The DFT of dimensions $N=8$ and $N=9$ have a maximal affine Hadamard family, where $d(F_8)=5$ and $d(F_9)=4$. Their structure can be consulted in Ref.~\cite{Tadej2006}. For $N=8$ there are $11117$ connected unlabeled graphs and we computed Eq.~\eqref{eq:f_numerical=0} for all of them. We obtained the $8$-dimensional QFTI from all graphs with a number of edges $|E|$ in the range $18\leq |E|\leq 28$ and in cases with fewer edges, but not in all of them. We checked the defect and the corresponding structure of the affine family in every case, counting a total of $2595$ solutions. The graphs with lower edges had $\lvert E \rvert=  12$, and interestingly, two of them were $3$-regular (see Fig.~\ref{fig:regular}).

For $N=9$ we did not study all graphs thoroughly, as there are $261080$ graphs, and solving Eq.~\eqref{eq:f_numerical=0} turned out computationally demanding as a consequence of having more variables to handle. We found solutions of Eq.~\eqref{eq:f_numerical=0} coinciding with the affine family of $F_9$ from all graphs with edges $24\leq |E|\leq36$ and with fewer edges as well. Interestingly, a $4$-regular graph ($\lvert E\lvert = 18$) appeared among the solutions below the $24$ edges (see Fig.~\ref{fig:regular}).

\subsection{\label{sec:qfti_N=6} Dimension six: non-affine case}

\begin{figure}[ht!]
    \centering
    \includegraphics[scale=0.15]{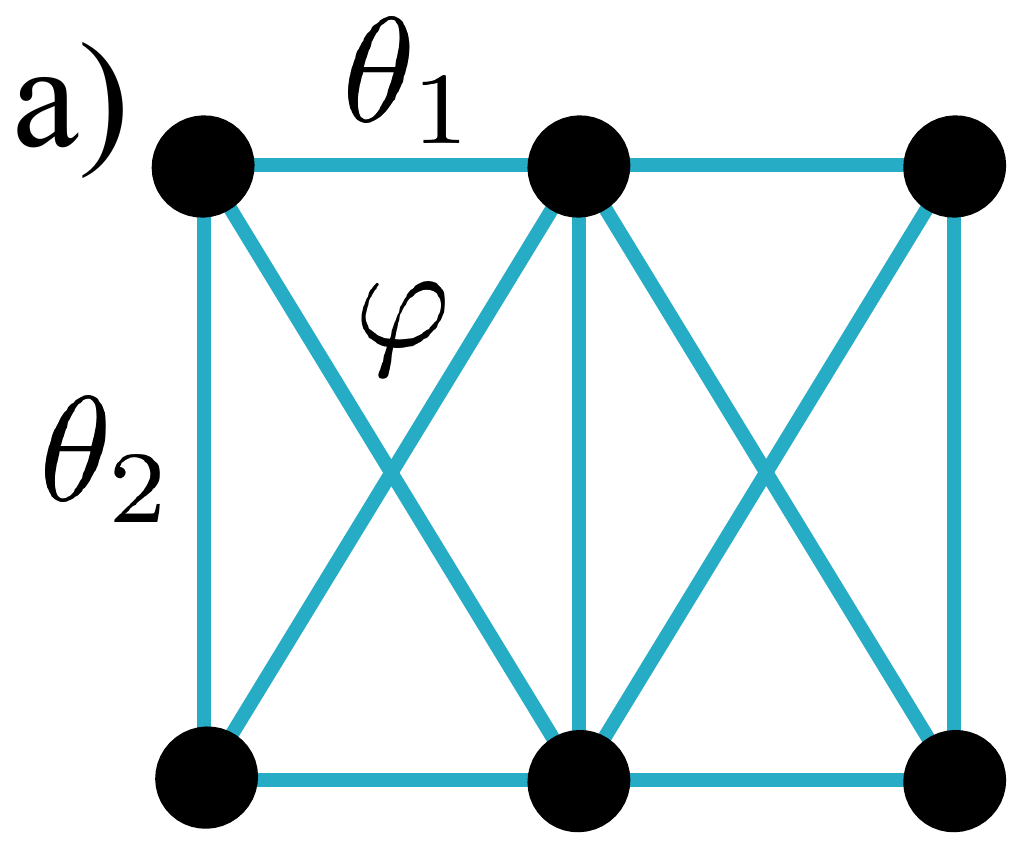}
    \includegraphics[width=\linewidth]{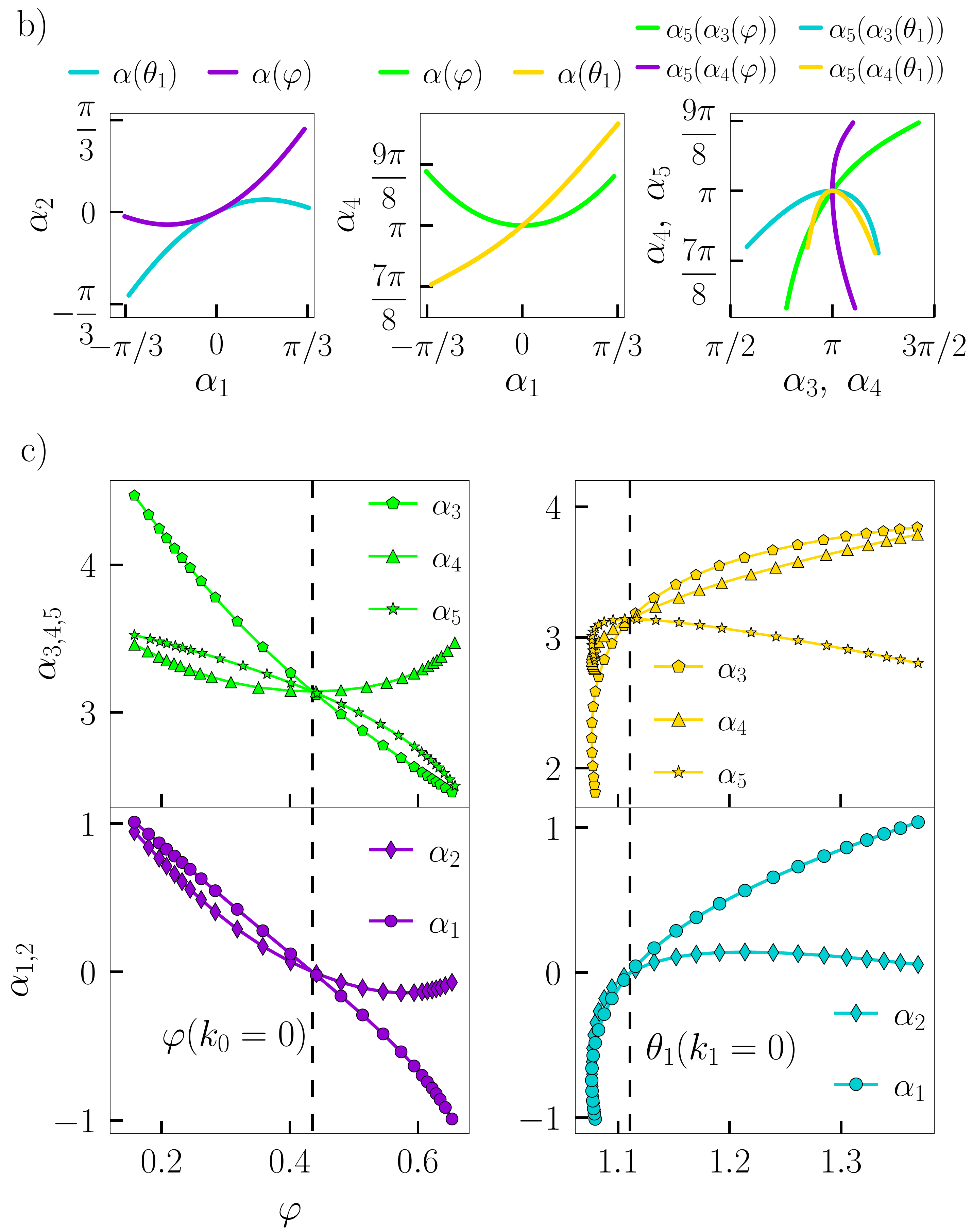}
    \caption{Free phases $\alpha_j$ present in the fourth column of Eq.~\eqref{eq:family_dft6}. b) Interdependence among the free phases $\alpha_j$ obtained from the geometrical configuration shown in a) according to the procedure described in Sec.~\ref{sec:qfti_N=6}. c) $\alpha_j$ ($j=1,2,3,4,5$) as function of the couplings $\varphi$ and $\theta_1$. The vertical dashed lines correspond to the solution for $\varphi$ and $\theta_1$ soved in Eq.~\eqref{eq:36_couplings}. The curves in b) are correlated with the curves in a) according to the Colors used on them.}
    \label{fig:free_phases_N=6}
\end{figure}

In this subsection, we comprehensively analyze the case of $N=6$. We show that the missing parameters predicted by the defect correspond to non-affine phases. We present a numerical procedure for identifying $F_6$ from the evolution operator $U$ and the difficulties in continuing to search for the $N$-dimensional QFTI for $N\geq 10$.

The CHMs in dimension $N=6$ are of great interest, as six is the first product of two distinct primes. A complete description of all $6\times6$ CHMs is still an open problem, despite all efforts in this field~\cite{szollosi2012}. Our interest is centered on $F_6$, whose defect is $4$, but its affine family only has the free phases $\phi_1$ and $\phi_2$, and reads
\begin{align}\label{eq:family_dft6}
    F_6 &=\frac{1}{\sqrt{6}}
    \begin{pmatrix}
        1 & 1 & 1 & \hphantom{-}1 & 1 & \hphantom{-}1\\
        1 & we^{i\phi_1} & w^2e^{i\phi_2} & -1 & w^4e^{i\phi_1} & w^5e^{i\phi_2}\\
        1 & w^2 & w^4 & \hphantom{-}1 & w^2 & w^4\\
        1 & -e^{i\phi_1} & e^{i\phi_2} & -1 & e^{i\phi_1} & -e^{i\phi_2} \\
        1 & w^4 & w^2 & \hphantom{-}1 & w^4 & w^2\\
        1 & w^5e^{i\phi_1} & w^4e^{i\phi_2} & -1 & w^2e^{i\phi_1} & we^{i\phi_2}
    \end{pmatrix},\\
    \nonumber w &=e^{-i\pi/3}.
\end{align}

When we calculated Eq.~\eqref{eq:f_numerical=0}, we found $F_6$ in a few cases, but only for $\phi_{1,2}=0$, not for any other value of $\phi_{1,2}$. All solutions of Eq.~\eqref{eq:f_numerical=0} had defect $4$ or $0$ (Tao's spectral matrix $S_6^{(0)}$), and by examining the arguments of $U=\exp{(-i\mathcal{C})}$ we could recognize a similar structure of Eq.~\eqref{eq:family_dft6}. From this equation we can see that the third column, second and fourth rows in the core should have fixed phases; nonetheless, we realized that free phases ($\neq \phi_{1},\phi_2$) were also immersed in these elements. The anchor column is supposed to have fixed phases $0$ and $\pi$, and we could notice that a column in $U$ tended towards these values. For the relevance of this column in the numerical routine, we refer to it as the anchor column, and we will denote its free phases as $\alpha_j$.

From the $112$ possible coupling configurations~\cite{CVETKOVIC1984}, after solving Eq.~\eqref{eq:f_numerical=0} with and without propagation constants, the normalization condition was met in $73$ cases (hexagon not counted) and in all of them we could find the $6$-dimensional QFTI. The corresponding graphs where those with edges in the range $8\leq\lvert E\rvert\leq 14$ and to ensure that from them Eq.~\eqref{eq:f_numerical=0} converged to $F_6$ we did as follows: for each graph we studied all the possible combinations of $\beta_l$ until the balanced condition was met; in each case, we searched for a column in the core where two phases, say $\alpha_{1,2}$, were subjected to the constraint $|\alpha_{1,2}|<1$. The presence of this column was the key to determine whether $F_6$ was accessible from the coupling configuration under inspection, because it corresponds to the anchor column (arguments $0$ and $\pi$) and $\alpha_{1,2}$ describe the elements with phase $0$. The elements with phase $\pi$ also had free phases involved, say $\alpha_{3,4,5}$, and were different from $\alpha_{1,2}$. In Fig.~\ref{fig:free_phases_N=6} we present the numerical results of $\alpha_{j}$ as a function of $C_{jk}$ for the interferometer where two quarters are superposed at two waveguides (see Figure~\ref{fig:free_phases_N=6} a)), which admits an analytical solution (see Appendix~\ref{Appendix:DFT6})
\begin{subequations}\label{eq:36_couplings}
    \begin{align}
        \varphi(k_0) &= \pm\frac{1}{\sqrt{2}}\arcsin{\left(\frac{1}{\sqrt{3}}\right)}+\frac{k_0\pi}{\sqrt{2}},\\
        \theta_1(k_1) &= \pm\frac{\pi}{2\sqrt{2}}+\frac{k_1\pi}{\sqrt{2}},\\
        \theta_2(k_2) &= \pm\frac{\pi}{4}+k_2\pi,
    \end{align}
\end{subequations}
where $k_m\in \mathbb{Z}$ and $\theta_1$, $\theta_2$, $\varphi$ are the couplings indicated in Fig.~\ref{fig:free_phases_N=6} a).

The phases $\phi_1$ and $\phi_2$ are the only two possible affine phases for $F_6$
(see Eq.~\eqref{eq:family_dft6}). The two missing degrees of freedom have been elusive~\cite{szollosi2012,Barros2013}, but our results on $\alpha_j$ presented in Fig.~\ref{fig:free_phases_N=6} confirm that $F_6$ belongs to a non-affine family. The new free phases $\alpha_j$ correspond to non-linear deformations of the couplings in Eq.~\eqref{eq:36_couplings}, which breaks the spatial symmetries of the array (see Fig.~\ref{fig:free_phases_N=6} a)).

\begin{figure}[ht!]
    \centering
    \includegraphics[width=\linewidth]{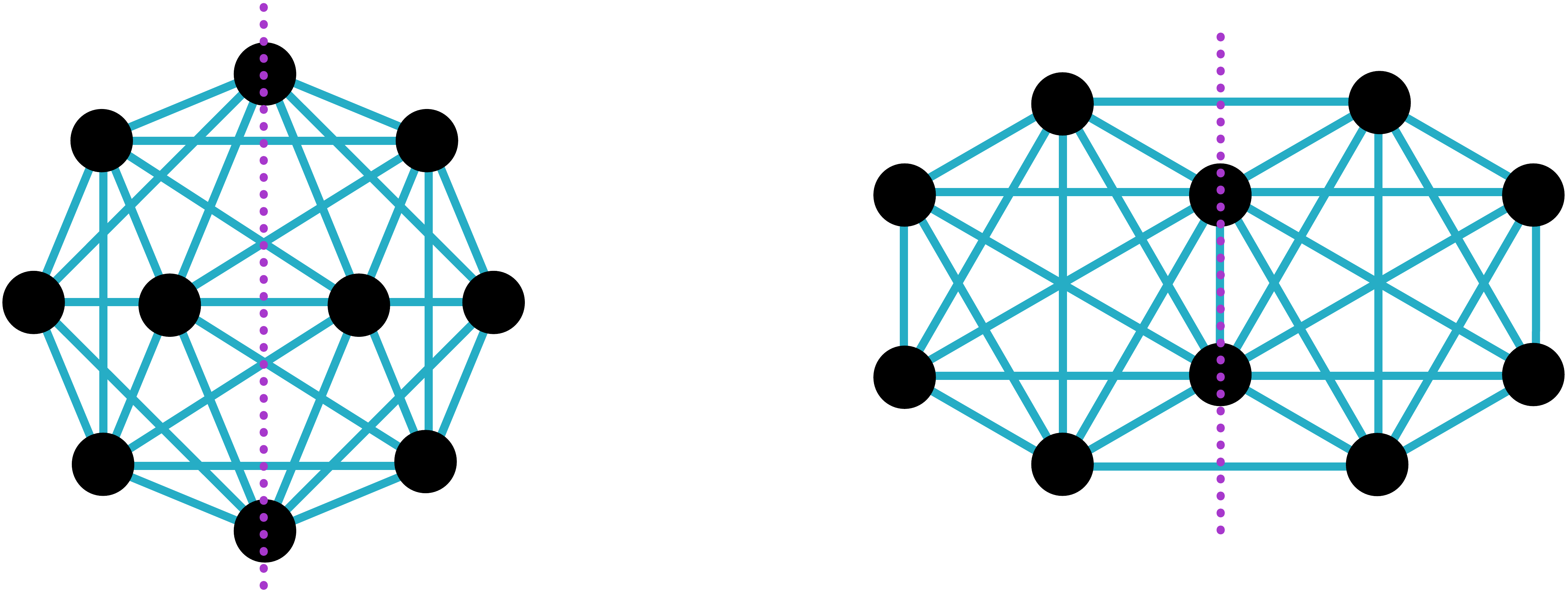}
    \caption{Two coupling configurations to implement the $10$-dimensional QFTI. The vertical dotted line indicates the axis of symmetry that the couplings must fulfill to be the QFTI.}
    \label{fig:dft10}
\end{figure}

For $N=8,\ 9$, the affine family encompasses the degrees of freedom determined by the defect, while for $N=10$ we have $d(F_{10})=8$, and the maximal affine family only contains four parameters. Following the same procedure explained for $N=6$, we endeavored to identify $F_{10}$ in the calculations, but did not continue because to find a solution of Eq.~\eqref{eq:f_numerical=0} in a reasonable time, the symmetries of $C_{jk}$ and $\beta_l$
must be enforced in the routine. This task implies that we first need a graphical representation of the graphs, because the graphs in our case are the spatial representation of the interaction in the interferometer. Even if we do this, there can be more than one spatial symmetry and they are intimately linked whether we can find CHMs or the DFT~\eqref{eq:dft}. In Figure~\ref{fig:dft10} we show two geometrical configurations for the $10$-dimensional QFTI with their corresponding axis of symmetry that yield $F_{10}$. 
Even for the decagon, where from the polygonal model we knew $F_{10}$ was reachable, solutions were hard to obtain without imposing the circulant constraint in the couplings. CHMs with defect $4$ were the main case we came across. In a forthcoming publication, we will address the CHMs that we found through this research and their geometrical configurations.

As there are too many graphs, approximately $\approx 10^6$ for $N=10$, performing the procedure we have described so far is impractical for every one of the graphs. An exploration by brute force is not suitable for $N\geq 10$.

\subsection{\label{sec:conjectures} Observations and conjectures to continue}

In addition to the details we mentioned to solve Eq.~\eqref{eq:f_numerical=0} and correctly procure the argument of $U_{jk}$, the number of graphs to inspect grows combinatorially with $N$, approximately $10^9$ for $N=11$, making an exhaustive analysis of all graphs computationally prohibitive. To guide the numerical efforts, we gather general observations of the graphs that are candidates for the $N$-dimensional QFTI.

Upon examination of the coupling configurations we found for $N=4,\ 5,\ 6,\ 7,\ 8,\ 9$, we observed three patterns to characterize the graphs behind $F_N$:\\
\textbullet\ First,
we conjecture that for prime $N$ all graphs with a number of edges $\lvert E\rvert$ in the range 
\begin{align}\label{eq:range_edges1}
    \lvert E(K_{N-1})\rvert\leq\lvert E\rvert\leq \lvert E(K_N)\rvert,
\end{align}
where $\lvert E(K_{N})\rvert=N(N-1)/2$, leads to the DFT.\\
\textbullet\ Second, for composite $N$, we conjecture that the lower bound of Eq.~\eqref{eq:range_edges1} falls below $\lvert E(K_{N-1})\rvert$ according to 
\begin{align}\label{eq:range_edges2}
    \lvert E(K_{N-1})\rvert-\ell \leq \lvert E\rvert \leq \lvert E(K_{N})\rvert,
\end{align}
where $\ell=2,\ 3,\ 4,\ldots$, starting with $\ell=2$ for $N=6$. The DFT arises from all graphs with edges in the range of Eq.~\eqref{eq:range_edges2}. \\
\textbullet\ Third, all the other possible solutions for non-prime $N$ are in the following range of edges
\begin{align}\label{eq:range_regular}
\lvert E(R_{\ell})\rvert \leq \lvert E\rvert < \lvert E(K_{N-1})\rvert-\ell,
\end{align}
where $R_{\ell}$ denotes $\ell$-regular graphs (same number of edges per vertex) and $\lvert E(R_{\ell})\rvert =N\ell/2$. However, in this case, not all graphs with edges in the range of Eq.~\eqref{eq:range_regular} are candidates for the $N$-dimensional QFTI, some cases do not even satisfy Eq.~\eqref{eq:f_numerical=0}.

The first conjecture comes from the observation of the number of variables involved in the solution of Eq.~\eqref{eq:f_numerical=0}. From Figure~\ref{fig:M5_all} we can see that, in general, as the number of couplings decreases, more propagation constants are involved, keeping the total number of variables fixed. The same happens for $N=7$. Consequently, the system of equations $|U_{jk}|^2-1/N$ and variables $C_{jk}$, $\beta_l$ is a square system of non-linear equations when the graphs from which we construct $\mathcal{C}$ are in the range of edges indicated in Eq.~\eqref{eq:range_edges1}. As the number of variables coincides with the number of equations, from an algebraic point of view it is plausible that this system of equations has solution.

\begin{table}[ht!]
\caption{\label{tab:conjecture} $k$-regular graphs (see Fig.~\ref{fig:regular}), and upper and lower bound for the range of number of edges of graphs leading to solutions for the $N$-dimensional QFTI.} 
\begin{ruledtabular}
\begin{tabular}{ccccc}
 $N$ & $\lvert E(K_N)\rvert$ & $\lvert E(K_{N-1})\rvert$ & $\lvert E\rvert$ & $k$-regular \\
\hline
$4$ &  $6$ &  $3$ &      &      \\
$5$ & $10$ &  $6$ &      &  $2$ \\
$6$ & $15$ &      &  $8$ &      \\
$7$ & $21$ & $15$ &      &  $4$ \\
$8$ & $28$ &      & $18$ &  $3$ \\
$9$ & $36$ &      & $24$ &  $4$ \\
$10$& $45$ &      &      &  $5$  
\end{tabular}
\end{ruledtabular}
\end{table}

The second conjecture is based on the solutions we found for $N=4,\ 6,\ 8,\ 9$ where all graphs with a given number of edges (Eq.~\eqref{eq:range_edges1} and Eq.~\eqref{eq:range_edges2}) led to $F_N$. From Table~\ref{tab:conjecture} we can see that when $N$ is not prime, the number of edges falls below $\lvert E(K_{N-1}) \rvert$. For $N=6$ instead of a lower bound of $\lvert E(K_5)\rvert =10$ edges, it is $\lvert E\rvert=8$. When we take the difference $\lvert E(K_{N-1})\rvert -\lvert E\rvert$ we notice that for $N=6$ the difference is $2$, for $N=8$ it is $3$ and for $N=9$ it is $4$. This strongly suggests that there should be a $\ell$-regular graph for $N=10$ as well, and it should be $\ell=5$. From the $60$ $5$-regular graphs for $N=10$, we obtained the $10$-dimensional QFTI in four of them (see Fig.~\ref{fig:regular}). We think that this trend continues, since the difference $\lvert E(K_{N-1}) \rvert-\lvert E \rvert = \ell$ is also the index of the $\ell$-regular graph below $\lvert E(K_{N-1})\rvert-\ell$ (except for $N=6$). All other solutions we found for non-prime $N$ live between the graphs with $\lvert E(K_{N-1})\rvert-\ell$ edges and the $\ell$-regular graphs, and this is the third conjecture.

The $\ell$-regular graphs are also present for prime $N$. For $N=5$ the cycle graph is the only array with five couplings where we obtained $F_5$ and is $2$-regular (see Fig.~\ref{fig:M5_all}). For $N=7$ a $4$-regular graph (see Fig.~\ref{fig:regular}) was also part of the graphs that converged to $F_7$  and has $14$ edges. We hypothesize that $2\ell$-regular graphs are candidates to implement the $N$-dimensional QFTI for prime $N$ with fewer than $\lvert E(K_{N-1})\rvert$ edges.

\subsection{\label{sec:tolerance} Physical admissibility and fabrication tolerance}

\begin{figure*}[t!]
    \centering
    \includegraphics[width=\linewidth]{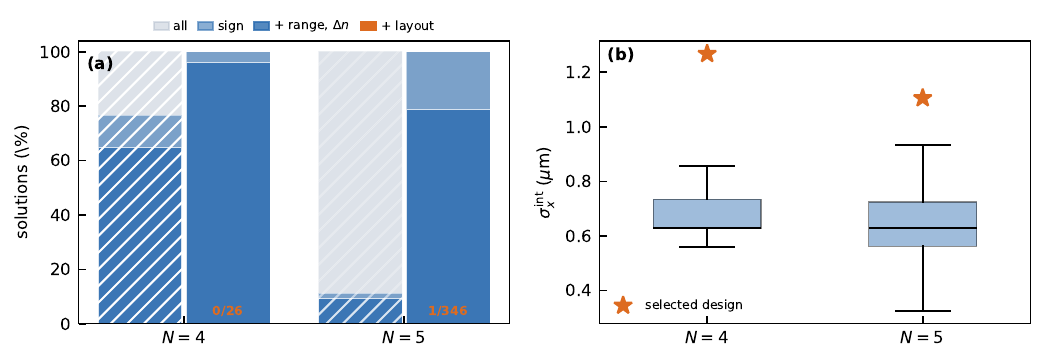}
    \caption{a) Fraction of solutions of Eq.~\eqref{eq:f_numerical=0} surviving each admissibility condition, for an unconstrained solver (hatched) and for a solver with positive couplings and bounded dynamic range (solid); the label gives the number of designs left after the layout test. b) Intrinsic positioning tolerance of the admissible solutions at fidelity $0.99$, that is, the error absorbed by the sensitivity of Eq.~\eqref{eq:infidelity} alone, excluding the systematic residual of the layout; the star marks the analytical design of Sec.~\ref{Sec:complete_graph}, measured with the same definition.}
    \label{fig:tolerance_screen}
\end{figure*}

\begin{figure*}[t!]
    \centering
    \includegraphics[width=\linewidth]{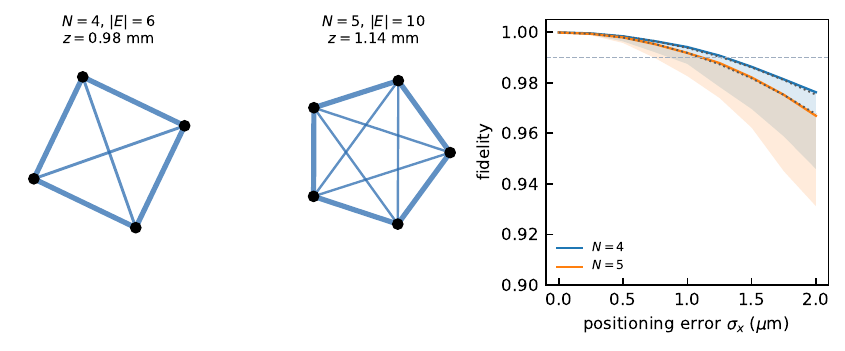}
    \caption{Selected coupling configurations. Cross sections are drawn to scale from the coordinates returned by the layout test, with line thickness proportional to the coupling strength. Right: fidelity against positioning error, with the shaded band covering the fifth percentile of $600$ realizations and the dotted line the prediction of Eq.~\eqref{eq:infidelity}.}
    \label{fig:tolerance_designs}
\end{figure*}

Since Equation~\eqref{eq:f_numerical=0} is solved without reference to the geometry that must host the couplings, its solutions are numerous. Before any of them reaches the laboratory two questions must be settled: whether a given coupling matrix corresponds to a realizable waveguide arrangement, and how sensitive the resulting QFTI is to fabrication error. Both are selection problems, and we address them with a criterion cheap enough to be applied to the complete solution set.

\textit{Error model.} Two mechanisms dominate. The transverse position of a waveguide enters as a \emph{relative} coupling error, $\delta C_{jk}/C_{jk}=\gamma\,\delta x$, because of the exponential decay of the evanescent coupling; the local index of refraction enters as an \emph{absolute} detuning, $\delta\beta_l=k_0\,\delta n_{\rm eff}$, and therefore perturbs every waveguide, including those with $\beta_l=0$. We use $\gamma=0.1\ \mu{\rm m}^{-1}$ and $C_{\max}=0.8\ {\rm mm^{-1}}$ from Sec.~\ref{sec:polygonal_model}, a minimum center-to-center separation $d_0=14\ \mu{\rm m}$, and $\sigma_n=5\times10^{-6}$ at $\lambda=1.55\ \mu{\rm m}$.

\textit{Sensitivity.} At a solution the fidelity $\mathcal{F}=\lvert{\rm tr}(U^{\dagger}U_{\delta})\rvert/N$ is maximal, its gradient vanishes, and the leading contribution is quadratic in the errors,
\begin{align}\label{eq:infidelity}
    1-\mathcal{F} &= \tfrac{1}{2}\left(\sigma_{\rm rel}^{2}\, S_C
    + \sigma_{\rm abs}^{2}\, S_\beta \right) + \mathcal{O}(\sigma^{3}),
\end{align}
where $S_C=\sum_{\langle j,k\rangle}M_{jk}$ runs over the edges of the graph, $S_\beta=\sum_{l}M_{l}$ over its vertices, and
\begin{align}\label{eq:metric}
    M_a &= \left\langle G_a^{\dagger}G_a\right\rangle
    -\left\lvert\left\langle G_a\right\rangle\right\rvert^{2},\qquad
    G_a=U^{\dagger}\partial_a U,
\end{align}
with $\langle X\rangle={\rm tr}(X)/N$ and $\partial_a U$ the Fr\'echet derivative of the matrix exponential with respect to the corresponding coupling or propagation constant. Each solution is therefore characterized by two numbers, obtained from one derivative per parameter and no sampling, at a cost of milliseconds. For the hexagon of Sec.~\ref{sec:polygonal_model}, Eq.~\eqref{eq:infidelity} reproduces a Monte Carlo simulation within $1\%$ up to $\sigma_{\rm rel}=10\%$, and $S_C=2.05$.

\textit{Admissibility.} A solution is retained when three conditions hold. (i)~The couplings admit a simultaneously positive gauge: a solution with mixed signs is realizable only if the signed graph is balanced, in which case $D={\rm diag}(\pm1)$ removes the signs and is absorbed by the phase shifters. (ii)~The ratio $C_{\max}/C_{\min}$ corresponds to an accessible spread of distances, $\Delta x=\gamma^{-1}\ln(C_{\max}/C_{\min})$, and the propagation constants stay within an accessible index contrast, $\Delta n_{\rm eff}\leq10^{-3}$. (iii)~A planar cross section exists: we assign target distances $d_{jk}=d_{0}+\gamma^{-1}\ln(C_{\max}/C_{jk})$ to the edges, require the remaining pairs to lie beyond the distance at which the coupling falls below one tenth of the weakest edge, and minimize the resulting stress over the $2N$ transverse coordinates. The residual $\delta d$ is a systematic error and enters the budget of Eq.~\eqref{eq:infidelity} as $\tfrac{1}{2}(\gamma\,\delta d)^{2}S_C$.

\textit{Selection.} Figure~\ref{fig:tolerance_screen}~a) reports the fraction of solutions surviving each condition, over a census of solutions of Eq.~\eqref{eq:f_numerical=0} obtained from random restarts on every connected graph. Solutions returned by an unconstrained root finder are frequently inadmissible, and increasingly so with $N$: the fraction admitting a positive gauge is $76\%$ for $N=4$ and $11\%$ for $N=5$. Enforcing $C_{jk}>0$ inside the solver removes the sign obstruction at no cost in graph coverage, which is $14$ of $21$ graphs for $N=5$ either way and $6$ of $6$ instead of $4$ of $6$ for $N=4$. The remaining conditions are what select: the index noise contributes an infidelity floor of $\sim10^{-4}$, two orders of magnitude below the target, so the dominant tolerance is positioning, and the intrinsic tolerance $\sigma_x^{\rm int}$ at $\mathcal{F}=0.99$ spans a factor of two across the admissible set [Fig.~\ref{fig:tolerance_screen}~b)]. The multiplicity of solutions is therefore a design resource rather than a redundancy: configurations implementing the same $F_N$ differ by that factor in how much fabrication error they absorb.

Applying the criterion to the analytical solutions of Sec.~\ref{Sec:complete_graph} selects the designs of Fig.~\ref{fig:tolerance_designs}. The quarter of Sec.~\ref{subsec:2D_3D}, with $\theta=\pi/4$, $\varphi=\pi/8$ and $\beta_l=0$, is realized as a square cross section of side $14.4\ \mu{\rm m}$ with an interaction length $z=0.98\ {\rm mm}$, and has $\sigma_x^{\rm int}=1.27\ \mu{\rm m}$. The pentagon of Sec.~\ref{sec:polygonal_model}, with the couplings of Eq.~\eqref{eq:5x5_solutions} for $k_1=1$, $k_2=0$, $n=1$, has a side of $14.4\ \mu{\rm m}$, a diagonal of $23.4\ \mu{\rm m}$, $z=1.14\ {\rm mm}$ and $\sigma_x^{\rm int}=1.11\ \mu{\rm m}$. Charging the systematic residual of the layout to the budget reduces these to $1.18\ \mu{\rm m}$ and $1.04\ \mu{\rm m}$, still comfortably above the sub-micron accuracy of femtosecond laser writing, so the two smallest QFTIs of this work are directly fabricable.

\section{\label{sec:FFT} Fast Fourier transform designs}

In this section, we address how our solutions can be used as larger building blocks to assemble the QFTI following the mathematical decomposition of the Fourier matrix through tensor products. We discuss the role of the propagation constants in determining what building blocks we can combine in the QFTI design. Then we present the scaling law for the spatial complexity~\cite{Li2025} to implement the QFTI based in our new building blocks.

When $N$ is a composite integer, the matrix factorization developed for the fast Fourier transform can be mapped to the design of optical~circuits~\cite{Barak2007}. In this regard, Crespi \textit{et al}.~\cite{Crespi2016} managed to experimentally build the circuit for $N=4$ and  $N=8$ following the formula from the Cooley-Tukey algorithm when $N=N_1N_2$
\begin{align}\label{eq:cooley-tukey}
    F_{N_1 N_2} &= P \left( \mathbb{I}_{N_1}\otimes F_{N_2}\right)T\left(F_{N_1}\otimes \mathbb{I}_{N_2}\right),
\end{align}
where $\otimes$ indicates the tensor product, $P$ is a permutation matrix, $\mathbb{I}_j$ is the identity, $F_{N_j}$ is the DFT and $T$ is a complex diagonal matrix known as the twiddle factor, and in our case it represents the action of the phase shifters. General details on permutations and twiddle factors can be consulted in Ref.~\cite{Tolimieri1997}. Eq.~\eqref{eq:cooley-tukey} is always valid, but if $N_1$ and $N_2$ are coprime, then the Good-Thomas algorithm
\begin{align}\label{eq:good-thomas}
    F_N &= P_1 \left(F_{N_1}\otimes \mathbb{I}_{N_2}  \right)\left(\mathbb{I}_{N_1}\otimes F_{N_2} \right) P_2,
\end{align}
provides a better decomposition, since it does not require twiddle factors, reducing the use of phase shifters.

\begin{figure}[ht!]
    \centering
    \includegraphics[width=1\linewidth]{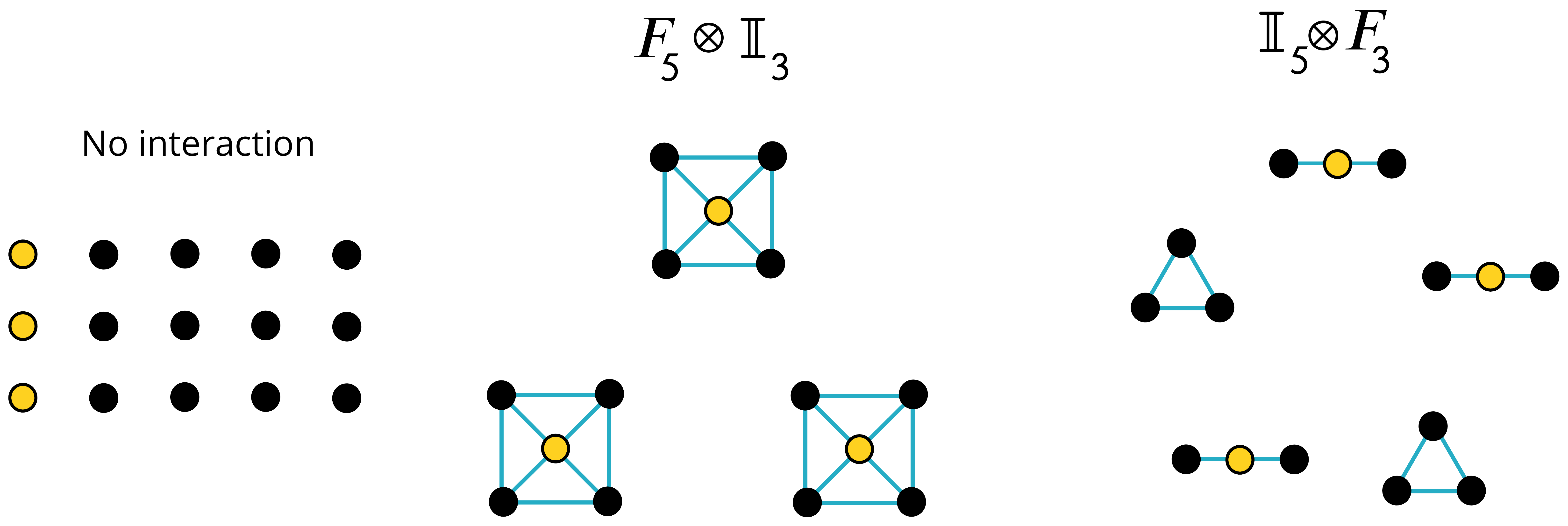}
    \caption{Implementation of the Good-Thomas algorithm in the design of the $15$-dimensional QFTI.}
    \label{fig:good-thomas}
\end{figure}

To assemble $F_N$ through Eqs.~\eqref{eq:cooley-tukey} and \eqref{eq:good-thomas} an important detail to be aware of is the presence of different $\beta_l$ in our building blocks $F_{N_j}$. In Eq.~\eqref{eq:U=exp_C} we assumed constant $C_{jk}$ and $\beta_l$, hence if we modify $\beta_l$ in one waveguide, its value must remain constant throughout the evolution. For example, to implement the $15$-dimensional QFTI we need the interferometers for $F_3$ and $F_5$. Let us take as the building block for the $5$-dimensional QFTI the case $2)$ in Fig.~\ref{fig:M5_all}, which requires one waveguide with $\beta\neq 0$. From the Good-Thomas decomposition~\eqref{eq:good-thomas}, in step $F_5\otimes\mathbb{I}_3$ the 15 waveguides will be group in three sets of five waveguides, performing three $F_5$ simultaneously (see Fig.~\ref{fig:good-thomas} b)), then in step $\mathbb{I}_{5}\otimes F_3$, five $F_3$ must be executed, but since we have three waveguides with $\beta\neq 0$, the tritter is no longer the valid interferometer in all five $F_3$. The tritter and the linear coupler must be combined in the last step (see Fig.~\ref{fig:good-thomas} c)).

The decomposition of $F_{N_1 N_2}$ through Eqs.~\eqref{eq:cooley-tukey} and~\eqref{eq:good-thomas} requires that we implement $N_1$ times $F_{N_2}$ and $N_2$ times $F_{N_1}$, and the same decomposition subsequently if $N_{1,2}$ is factorable. In this way, the total number of interferometers is captured by the arithmetic derivative $D(N_1 \cdot N_2) = N_1 D(N_2) + N_2 D(N_1)$. Having $N=N_1N_2\cdots N_q$, by the generalized Leibniz rule we have
\begin{subequations}
    \begin{align}
        D(N_1 \cdot N_2 \cdots N_q) &= \sum\limits_{j=1}^q \left( D(N_j) \prod\limits_{k \neq j} N_k \right),\\
        D(N) &= N \sum\limits_{j=1}^{q} \frac{D(N_j)}{N_j},
        \label{eq:total_interferometers}
    \end{align}
\end{subequations}
where we have used the fact that the product of all $N_k$ for $k\neq j$ is $N/N_j$. If we take $D(N_j)=1$ when for $N_j$ we know the geometrical configuration for the $N_j$-dimensional QFTI, then $D(N)$~\eqref{eq:total_interferometers} is the total number of interferometers that we will need to implement $F_{N}$. Notice that $D(N)$, as we have defined, is multivalued and depends on the particular choice of factors $N_j$.

The design proposed by Barak \textit{et al.}~\cite{Barak2007} is based on Eq.~\eqref{eq:cooley-tukey}. It was meant to be used to implement the quantum Fourier transform (QFT) with linear optics and rely on $2\times 2$ interferometers. For the $n$-qubit QFT we would need $D(N=2^n)=n2^{n}/2$ interferometers. With our building blocks for $N=4$ and  $N=8$, this number can be further reduced. Using $n_1=4$ and $n_2=8$ in Eq.~\eqref{eq:total_interferometers} we have
\begin{subequations}
    \begin{align}
        D_{4,8} &= 4^{k_1}8^{k_2}\left(\sum\limits_{j=1}^{k_1}\frac{1}{4}+\sum\limits_{j=1}^{k_2}\frac{1}{8}\right) ,\\
        D_{4,8} &= \frac{1}{4}\left(1-\frac{1}{\frac{3}{2}+\frac{k_1}{k_2}}\right)D_2,
        \label{eq:D_48}
    \end{align}
\end{subequations}
where $D_2$ is the number of $2\times 2$ interferometers that would be needed to implement the $(2k_1+3k_2)$-qubit QFT, that is, $D_2=(2k_1+3k_2)2^{2k_1+3k_2}/2$.

In practice, Eq.~\eqref{eq:D_48} tells us that by combining the coupling configurations we have found for $N=4$ and  $N=8$, for example, we would only need a total of $768$ interferometers to implement the $10$-qubit QFT or equivalently the $1024$-dimensional QFTI. Based on Reck's or Clements' architectures, we would need approximately $5\times 10^{5}$ interferometers, which is beyond current expectations in mid-term on the scaling in photonic hardware~\cite{Lopez2025}.

\begin{figure}[ht!]
    \centering
    \includegraphics[width=1\linewidth]{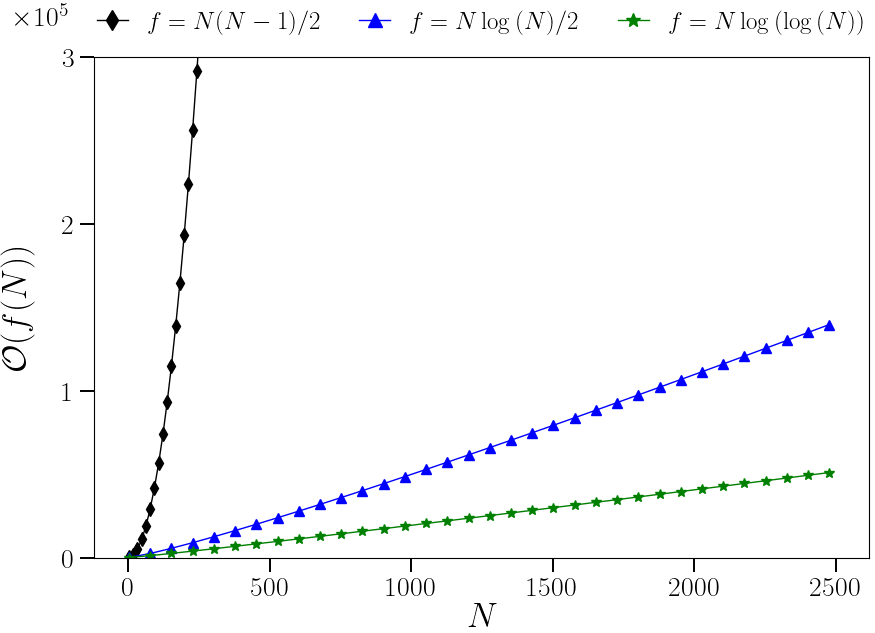}
    \caption{Spatial complexities of Reck, Clements and Barak architectures and the new scaling law based on the decomposition through Eq.~\eqref{eq:cooley-tukey} and Eq.~\eqref{eq:good-thomas} of building blocks for the $N$-dimensional QFTI for $N=2,\ldots,10$. }
    \label{fig:scaling_law}
\end{figure}

For the assessment of boson sampling devices, we need a multiarm interferometer that implements at least the $2500$-dimensional QFTI~\cite{Clifford2018}. From our building blocks for $N=4,\ 5$ a total of $2625$ interferometers would be necessary, in contrast to the approximately $3\times10^6$  of $2\times2$ beamsplitter in Reck's and Clements' architectures. This can be improved if we construct the $2520$-dimensional QFTI ($9\times8\times7\times5$) instead, and, since the factors of $2520$ are coprimes, by means of the Good-Thomas decomposition~\eqref{eq:good-thomas} we would require just $1459$ interferometers and less phase-shifters (no twiddle factors). As the number of interferometers $D(N)$~\eqref{eq:total_interferometers} depends on the size of the building blocks, the scaling law of Eq.~\eqref{eq:total_interferometers} is governed by the number of factors of $N$. From the Hardy–Ramanujan theorem~\cite{Ramanujan} we know that the number of distinct prime factors in $N$ are typically $\log{\log{N}}$, hence, we have $\mathcal{O}(N\log{\log{N}})$ for $D(N)$~\eqref{eq:total_interferometers} on average. In Figure~\ref{fig:scaling_law} we present the scaling laws for the number of interferometers to perform the $N$-dimensional QFTI based on the results of Reck-Clements, Barak and ours.

\section{\label{sec:conclusions} Outlook and conclusions}

We have demonstrated that the evolution operator in a coupled bosonic Hamiltonian can be tailored to be the $N$-dimensional QFTI in a single stage of evolution accompanied by input and output phase shifters. Our research provides two complementary frameworks: polygonal and graph-based models. The former admits analytical solutions for $N=4,\ 5,\ 6$ and a general well-defined theoretical formalism, which allowed us to find numerical solutions for the couplings up to $N=31$. Based on experimental data, we determined that this model is bound to $N\leq 6$, nevertheless Abarca-Ram\'irez \textit{et al}. recently showed that long-range evanescent coupling can be achieved through photonic molecules~\cite{Abarca_Ramirez2025}, opening up the possibility of having larger polygonal multiarm interferometers.

To overcome the limitation of the exponential decay of the evanescent coupling strength with distance present in the polygonal model, we introduce the graph-based model, where we explored the geometrical configurations according to the mapping between the connected unlabeled graphs and a system of evanescently coupled waveguides. In contrast to the polygonal model, different propagation constants are required in the Hamiltonian. 
We found all the geometrical configurations to implement the $N$-dimensional QFTI for $N\leq 8$ and an exhaustive but not complete exploration for $N=9$ and some cases for $N=10$.

We have provided the building blocks for the implementation of any QFTI of size $N=2^{n_1}3^{n_2} 5^{n_3} 7^{n_4}$ using the Cooley-Tukey and Good-Thomas decompositions. The experimental implementation of the PICs we are proposing requires meticulous examination of the numerical solutions for the couplings and propagation constants to check their feasibility in the laboratory.

We discussed the difficulties (symmetries and unknown free phases) in continuing the exploration of coupling configurations capable of performing $F_N$ for $N\geq 10 $, and, based on the numerical evidence, we presented three conjectures on where to find it. From our search for the $6$-dimensional QFTI we uncovered the missing non-affine parameters of $F_6$. Our findings are a strong support for the conjecture that there is a $4$-parameter family of CHMs, to which $F_6$ belongs~\cite{Bengtsson2007}. This result is relevant to the search for a fourth mutually unbiased basis of order $6$ because, within our theoretical framework, the missing degrees of freedom can finally be explored.

The new multiarm interferometers we have found to perform the DFT can be used as Hadamard gates for qudits in the KLM protocol~\cite{Karacsony2024}. They can also be used as mixing layer in universal layered programmable decompositions~\cite{Alvarez-Vizoso}. This takes special importance for boson sampling, where, to prove quantum computational advantage, the circuit must be capable of implementing any unitary operator.

\begin{acknowledgments}
The authors thank Dardo Goyeneche for useful conversations.
E.B., C.M., A. M., C. H.-A., and A.D. were supported by the National Agency of Research and Development (ANID) through  FONDECYT REGULAR No. 1230897 and No. 1240204, and by the Millennium Science Initiative Program -- ICN17$_-$012.

\end{acknowledgments}

\appendix

\appendix{\counterwithin{equation}{section}}
\appendix

\section{\label{Appendix:norm_angles} Balanced condition and arguments}

This appendix derives the two conditions that Eq.~\eqref{eq:f_numerical=0} encodes, for the polygonal model of Sec.~\ref{sec:polygonal_model}, where the coupling matrix $\mathcal{C}$ of Eq.~\eqref{eq:Coupling_polygon} is circulant. Both conditions are obtained in closed form: the balanced condition becomes an algebraic identity in the eigenvalues, and the condition on the arguments reduces to the statement that a single sequence of phases is quadratic in its index. The construction also yields the origin of the permutation matrices observed in Sec.~\ref{Sec:complete_graph}.

\subsection{The symbol of the evolution}

A real symmetric circulant matrix is diagonalized by the Fourier modes $(v_k)_j=e^{2\pi ijk/N}/\sqrt{N}$, with eigenvalues
\begin{align}\label{eq:appA_eigen}
    \lambda_k &= C_0+2\sum_{s=1}^{\lfloor N/2\rfloor} C_s
    \cos\!\left(\frac{2\pi sk}{N}\right),
\end{align}
where $C_0=\beta$ and $C_s=C_{N-s}$. Since $U=e^{-i\mathcal{C}}$ shares these eigenvectors, its entries depend only on the difference of indices,
\begin{align}\label{eq:appA_symbol}
    U_{mn} &= \frac{1}{N}\,S(m-n),  &
    S(r) &= \sum_{k=0}^{N-1} e^{2\pi ikr/N-i\lambda_k},
\end{align}
and we call $S$ the symbol of the evolution. Equation~\eqref{eq:appA_eigen} gives $\lambda_k=\lambda_{N-k}$, so $S(r)=S(-r)$ and the sign convention for $m-n$ in Eq.~\eqref{eq:appA_symbol} is immaterial.

That same degeneracy is what reduces the $N$ terms of $S$ to $\lfloor N/2\rfloor+1$. Pairing $k$ with $N-k$ and using $e^{2\pi ir}=1$ we obtain
\begin{align}\label{eq:appA_reduced}
    S(r) &= \sum_{k=0}^{M} A_k(r)\, e^{-i\lambda_k}, &
    M &= \left\lfloor \frac{N}{2}\right\rfloor,
\end{align}
where $A_0=1$ and $A_k(r)=2\cos(2\pi kr/N)$ for every paired index. For odd $N$ all $k=1,\dots,M$ are paired; for even $N=2M$ the index $k=M$ is its own partner and contributes $A_M(r)=(-1)^{r}$ instead.

A first consequence of Eq.~\eqref{eq:appA_reduced} is that the propagation constants are redundant in the polygonal model. Adding a constant to every $\lambda_k$, which is what $C_0=\beta$ does in Eq.~\eqref{eq:appA_eigen}, multiplies $S$ by a global phase, leaving both the modulus of $S$ and the dephased matrix unchanged. For a circulant array we may therefore set $\beta=0$ without loss of generality, as the solutions of Sec.~\ref{sec:polygonal_model} do.

\subsection{Balanced condition}

The balanced condition $|U_{mn}|^2=1/N$ reads $|S(r)|^2=N$ for every $r$. From Eq.~\eqref{eq:appA_reduced},
\begin{align}\label{eq:appA_modulus}
    |S(r)|^2 &= \sum_{k=0}^{M}A_k^2
    +2\!\!\sum_{0\leq k<l\leq M}\!\! A_kA_l\cos(\lambda_k-\lambda_l).
\end{align}
The first sum is independent of the eigenvalues and can be evaluated in closed form. Writing $A_k^2=2+2\cos(4\pi kr/N)$ for the paired indices and using
\begin{subequations}\label{eq:appA_sums}
    \begin{align}
    \sum_{k=1}^{M}\cos\!\left(\frac{4\pi kr}{N}\right)
    &= \frac{1}{2}\sum_{s=1}^{N-1}\cos\!\left(\frac{2\pi sr}{N}\right)
    = \frac{N\delta_{r0}-1}{2},\\
    \sum_{k=1}^{M-1}\cos\!\left(\frac{2\pi kr}{M}\right)
    &= M\,\delta_{r\equiv 0\, ({\rm mod}\, M)}-1,
    \end{align}
\end{subequations}
the first for odd $N$, where doubling the index permutes the residues, and the second for even $N=2M$, one finds
\begin{align}\label{eq:appA_diagonal}
    \sum_{k=0}^{M}A_k^2 &=
    \begin{cases}
        (N-1)+N\delta_{r0}, & N \text{ odd},\\[2pt]
        (N-2)+N(\delta_{r0}+\delta_{rM}), & N \text{ even}.
    \end{cases}
\end{align}
Substituting Eq.~\eqref{eq:appA_diagonal} into Eq.~\eqref{eq:appA_modulus} and splitting the pair sum according to whether an index equals $0$ or $M$ gives the identity
\begin{align}\label{eq:appA_identity}
    |S(r)|^{2} &= N + E(r),
\end{align}
valid for arbitrary $\lambda_k$, where for odd $N$
\begin{align}\label{eq_appe:Norm_odd}
    \nonumber E(r) &= 8\sum_{j=1}^{\frac{N-3}{2}}\;
    \sum_{k=j+1}^{\frac{N-1}{2}}
    \cos\!\left(\frac{2\pi jr}{N}\right)\cos\!\left(\frac{2\pi kr}{N}\right)
    \cos\left(\lambda_j-\lambda_k\right)\\
    &+4\sum_{j=1}^{\frac{N-1}{2}}\cos\!\left(\frac{2\pi jr}{N}\right)
    \cos\left(\lambda_0-\lambda_j\right)+N\delta_{r0}-1,
\end{align}
and for even $N=2M$
\begin{align}\label{eq_appe:Norm_even}
    \nonumber E(r) &=
    8\sum_{j=1}^{M-2}\sum_{k=j+1}^{M-1}
    \cos\!\left(\frac{\pi jr}{M}\right)\cos\!\left(\frac{\pi kr}{M}\right)
    \cos\left(\lambda_j-\lambda_k\right)\\
    \nonumber &+4\sum_{j=1}^{M-1}\cos\!\left(\frac{\pi jr}{M}\right)
    \left[\cos\left(\lambda_0-\lambda_j\right)
    +(-1)^{r}\cos\left(\lambda_j-\lambda_M\right)\right]\\
    &+2(-1)^{r}\cos\left(\lambda_M-\lambda_0\right)
    +N(\delta_{r0}+\delta_{rM})-2.
\end{align}
The array is balanced if and only if $E(r)=0$ for every $r$.

Two remarks fix the size of this system. First, $A_k(N-r)=A_k(r)$, so $E(N-r)=E(r)$ and only $r=0,1,\dots,M$ need be imposed. Second, unitarity gives the sum rule $\sum_{r}|S(r)|^{2}=N^{2}$, which is satisfied identically when all $|S(r)|^2=N$ and therefore renders one of those $M+1$ equations redundant. The balanced condition is thus a system of $\lfloor N/2\rfloor$ independent equations, as stated in Sec.~\ref{sec:polygonal_model}, and with $\beta=0$ its unknowns are the $\lfloor N/2\rfloor$ couplings $C_1,\dots,C_{M}$: the system is square. For $N=4$ it reproduces the two conditions of Eq.~\eqref{eq:4x4_normalization}, and for $N=5$, where $\lambda_k=2\theta\cos(2\pi k/5)+2\varphi\cos(4\pi k/5)$, the two equations $E(1)=E(2)=0$ reproduce Eq.~\eqref{eq:pentagon}. In the latter case the elimination of $\theta+\varphi$ through $\sin[5(\theta+\varphi)/2]=0$ leaves
\begin{align}\label{eq:appA_pentagon_fixed}
    0 &= 4\cos^{2}\!\left(\frac{\sqrt{5}}{2}(\theta-\varphi)\right)
    +2(-1)^{k_1}\cos\!\left(\frac{\sqrt{5}}{2}(\theta-\varphi)\right)-1,
\end{align}
whose roots are $\cos u=[-(-1)^{k_1}\pm\sqrt{5}]/4$. The factor $(-1)^{k_1}$ is what admits $n=2,\ 4$ for even $k_1$ in Eq.~\eqref{eq:5x5_solutions}; without it only the odd-$k_1$ branch $n=1,\ 3$ is recovered.

\subsection{Arguments}

Once the array is balanced we may write $S(r)=\sqrt{N}\,e^{i\sigma_r}$, so that all the information left in $U$ is the phase sequence $\sigma_r$. The dephased matrix of Eq.~\eqref{eq:U_dephased} has entries
\begin{align}\label{eq:appA_dephased}
    D_{mn} &= \frac{1}{\sqrt{N}}\,
    e^{i\left(\sigma_{m-n}-\sigma_{-n}-\sigma_{m}+\sigma_{0}\right)},
\end{align}
since dividing by the first row and the first column removes $\sigma_{m}$ and $\sigma_{-n}$ and restores the phase $\sigma_0$ of the corner. Requiring $D=F_N$, that is $D_{mn}=\omega^{mn}/\sqrt{N}$ with $\omega=e^{-2\pi i/N}$, therefore amounts to the single functional equation
\begin{align}\label{eq:appA_cocycle}
    \sigma_{m-n}-\sigma_{m}-\sigma_{-n}+\sigma_{0}
    &\equiv -\frac{2\pi mn}{N} \pmod{2\pi}.
\end{align}
This is the entire content of the condition on the arguments, and it replaces the expansion of a sum of four arctangents by a linear problem. Its general solution is quadratic. Inserting $\sigma_r=\sigma_0+\alpha r+b r^{2}$ into Eq.~\eqref{eq:appA_cocycle} gives $-2bmn\equiv-2\pi mn/N$, so
\begin{align}\label{eq:appA_quadratic}
    \sigma_r &= \sigma_0+\alpha r+\frac{a\pi r^{2}}{N},
    & a &\in \mathbb{Z},
\end{align}
and periodicity $\sigma_{r+N}\equiv\sigma_r$ forces $\alpha\equiv-a\pi+2\pi s/N$ with $s\in\mathbb{Z}$. The linear coefficient cancels identically in Eq.~\eqref{eq:appA_cocycle}, so it carries no information about the couplings; it is nevertheless fixed by the symmetry $S(r)=S(-r)$ of Eq.~\eqref{eq:appA_symbol}, which requires $2s\equiv 0$ $({\rm mod}\ N)$ and therefore $\alpha=-a\pi$ for odd $N$. The phase sequence of a circulant QFTI is then determined by a single integer,
\begin{align}\label{eq:appA_sigma}
    \sigma_r-\sigma_0 &= \frac{a\pi\, r(r-N)}{N},
\end{align}
so that the $\lfloor N/2\rfloor$ conditions on the arguments are not independent transcendental equations but one arithmetic pattern. With Eq.~\eqref{eq:appA_quadratic} the dephased matrix is
\begin{align}\label{eq:appA_result}
    D_{mn} &= \frac{1}{\sqrt{N}}\,\omega^{amn},
\end{align}
which is a discrete Fourier transform if and only if $\gcd(a,N)=1$, and equals $F_N$ after relabelling the columns as $n\mapsto a^{-1}n \bmod N$.

Equations~\eqref{eq:appA_quadratic} and~\eqref{eq:appA_result} give a compact criterion. A circulant array implements the $N$-dimensional QFTI when its symbol has constant modulus $\sqrt{N}$ and a phase that is quadratic in the index with curvature $a\pi/N$ for some $a$ coprime with $N$; the integer $a$ is a property of the solution, and the permutation matrices reported in Sec.~\ref{Sec:complete_graph} are exactly the relabelling $n\mapsto a^{-1}n$. No permutation is needed when $a=1$.

The two analytical solutions of Sec.~\ref{Sec:complete_graph} illustrate this. For the quarter with $\theta=\pi/4$ the curvature gives $a=1$ when $\varphi=\pi/8$ and $a=3\equiv-1$ when $\varphi=-\pi/8$; in the second case $a^{-1}=3$ and the relabelling $n\mapsto-n$ is the exchange of the second and fourth columns. For the pentagon of Eq.~\eqref{eq:5x5_solutions} with $k_2=0$ the curvature yields $a=1,2$ for odd $k_1$ with $n=1,3$ and $a=3,4$ for even $k_1$ with $n=2,4$. The case $k_1=2$, $n=2$ has $a=3$ and $a^{-1}=2$, so the columns must be relabelled as $n\mapsto 2n$, which is the permutation $P=[\hat e_1,\hat e_3,\hat e_5,\hat e_2,\hat e_4]$ quoted in Sec.~\ref{sec:polygonal_model}.

Finally, Eq.~\eqref{eq:appA_quadratic} identifies the role of the phase shifters. Bluestein's identity $2mn=m^{2}+n^{2}-(m-n)^{2}$ factors the Fourier kernel as
\begin{align}\label{eq:appA_bluestein}
    \omega^{mn} &= \omega^{m^{2}/2}\,\omega^{-(m-n)^{2}/2}\,\omega^{n^{2}/2},
\end{align}
a product of two diagonal chirps and one circulant chirp. A balanced circulant evolution whose symbol obeys Eq.~\eqref{eq:appA_quadratic} is precisely that circulant factor, because $\omega^{-r^{2}/2}=e^{i\pi r^{2}/N}$, and $\Phi^{\rm in}$ and $\Phi^{\rm out}$ supply the two diagonal chirps. The single-stage QFTI of this work is, in this sense, a physical realization of Bluestein's factorization in which the chirp convolution is performed by evanescent coupling rather than by a fast Fourier transform.

\section{6-dimensional DFT\label{Appendix:DFT6}}

By expanding Eq.~\eqref{eq:norm_equation} for $N=6$ we obtain the following set of equations
\begin{widetext}
\begin{subequations}\label{eq_appe:norm_6x6}
        \begin{align}
        0 &= 2 + 2\cos{(\lambda_1 - \lambda_0)} + 2\cos{(\lambda_2 - \lambda_0)} + \cos{(\lambda_3 - \lambda_0)} +2\cos{(\lambda_3-\lambda_1)} + 2\cos{(\lambda_3-\lambda_2)} + 4\cos{(\lambda_2-\lambda_1)}, \\
        0 &= 1 - \cos{(\lambda_1 - \lambda_0)} + \cos{(\lambda_2 - \lambda_0)} + \cos{(\lambda_3 - \lambda_0)}  + \cos{(\lambda_3 - \lambda_1)} - \cos{(\lambda_3 - \lambda_2)}+ \cos{(\lambda_2 - \lambda_1)}, \\
        0 &= 1 + \cos{(\lambda_1 - \lambda_0)} + \cos{(\lambda_2 - \lambda_0)} - \cos{(\lambda_3 - \lambda_0)} + \cos{(\lambda_3 - \lambda_1)} + \cos{(\lambda_3 - \lambda_2)} - \cos{(\lambda_2 - \lambda_1)}, \\
        0 &= 2 - 2\cos{(\lambda_1 - \lambda_0)} + 2\cos{(\lambda_2 - \lambda_0)} - \cos{(\lambda_3 - \lambda_0)} + 2\cos{(\lambda_3 - \lambda_1)}- 2\cos{(\lambda_3 - \lambda_2)}- 4\cos{(\lambda_2 - \lambda_1)},
    \end{align}
\end{subequations}
\end{widetext}
where, according to Eq.~\eqref{eq:lambda_even}, the eigenvalues $\lambda_j$ are given by
\begin{align}
    \lambda_0 &= 2C_1+2C_2+C_3,\\
    \lambda_1 &= C_1-C_2-C_3,\\
    \lambda_2 &= -C_1-C_2+C_3.
\end{align}

We want to find the coupling coefficients $C_1$, $C_2$ and $C_3$. To solve Eqs.~\eqref{eq_appe:norm_6x6}, we enforce the following conditions
\begin{subequations}\label{eq_appe:condition}
    \begin{align}
        \cos{(\lambda_2-\lambda_0)} &= -\frac{1}{2},\\
        \cos{(\lambda_3-\lambda_1)} &= -\frac{1}{2}.
    \end{align}
\end{subequations}

Replacing the conditions of Eqs.~\eqref{eq_appe:condition} in Eqs.~\eqref{eq_appe:norm_6x6}, we have
\begin{align}
    3(C_1+C_2) &=\frac{4\pi}{3}+2\pi k_1,\\
    3(C_1-C_2) &= \frac{2\pi}{3}+2\pi k_2,
\end{align}
which leads to $C_1=\pi/3$ and $C_2=\pi/9$. This implies that
\begin{align}
    \cos{\left(\lambda_3-\lambda_2+\lambda_1-\lambda_0\right)} &= 0,\\
    \cos{(2C_3+C_1)} &= 0,\\
    2C_3+C_1&=\frac{\pi}{2}+k_3\pi,\\
    C_3 &= \frac{\pi}{12}.
\end{align}

Taking into account the periodicity of trigonometric solutions, the full set of soluionts for Eq.~\eqref{eq_appe:norm_6x6} is finally given by
\begin{subequations}\label{eq:N=6_solutions}
    \begin{align}
        C_1 &= \frac{\pi}{3}+\frac{\pi}{3}( k_1+k_2),\\
        C_2 &= \frac{\pi}{9}+\frac{\pi}{3}(k_1-k_2),\\
        C_3 &=\frac{\pi}{12}-\frac{\pi}{6}( k_1+k_2-3k_3).
    \end{align}
\end{subequations}

\begin{figure}
    \centering
    \includegraphics[width=0.35\linewidth]{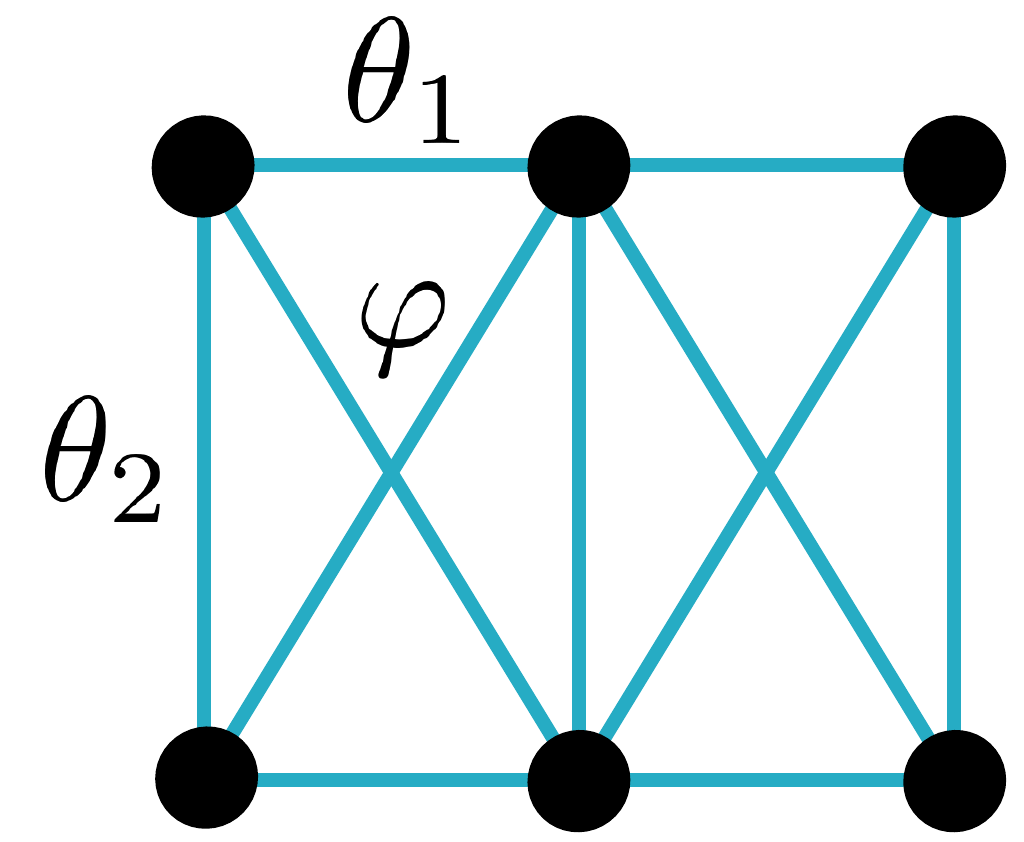}
    \caption{Cross section for one of the $74$ coupling configurations where a QFTI is possible. Black circles: waveguides; blue lines: couplings.}
    \label{fig:qfti6}
\end{figure}

For the array shown in Fig.~\ref{fig:qfti6} the coupling matrix reads
\begin{align}
    \label{eq_appe:C_6x6}\mathcal{C} &=
    \begin{pmatrix}
        0 & \theta_2 & \theta_1 & 0 & 0 & \varphi \\
        \theta_2 & 0 & \varphi & 0 & 0&\theta_1 \\
         \theta_1 &\varphi & 0 & \varphi & \theta_1 & \theta_2 \\
        0 & 0 & \varphi & 0 & \theta_2 & \theta_1 \\
        0 & 0 & \theta_1 &  \theta_2 & 0 & \varphi \\
        \varphi & \theta_1 & \theta_2 & \theta_1 & \varphi & 0
    \end{pmatrix}.
\end{align}

Notice that Eq.~\eqref{eq_appe:C_6x6} has the block form
\begin{align}
    \mathcal{C} &=
    \begin{pmatrix}
        A & B \\
        B & A
    \end{pmatrix},\\
    A&=\begin{pmatrix}
        0 & \theta_2 & \theta_1\\
        \theta_2 & 0 & \varphi\\
        \theta_1 & \varphi & 0
    \end{pmatrix},\\ B &=
    \begin{pmatrix}
        0 & 0 & \varphi\\
        0 & 0 & \theta_1\\
        \varphi & \theta_1 & \theta_2
    \end{pmatrix}.
\end{align}
Using the orthogonal matrix $M$
\begin{align}
    M &=\frac{1}{\sqrt{2}}
    \begin{pmatrix}
        \mathbb{I}_3 &\hphantom{-} \mathbb{I}_3\\
        \mathbb{I}_3 & -\mathbb{I}_3
    \end{pmatrix},
\end{align}
where $\mathbb{I}_3$ is the identity matrix, a block diagonalization of $\mathcal{C}$ is possible and results in
\begin{align}\label{eq_appe:block_diagonal}
    M \mathcal{C} M &= 
    \begin{pmatrix}
        A + B & 0 \\
        0 & A-B
    \end{pmatrix}
\end{align}

Using Eq.\eqref{eq_appe:block_diagonal}, we can calculate the complex exponential of $\mathcal{C}$ as
\begin{align}
    \nonumber e^{-i\mathcal{C}}\! &=\! M\!
    \begin{pmatrix}
        e^{-i(A+B)} & 0\\
        0 & e^{-i(A-B)}
    \end{pmatrix}\! M,\\
    \!&=\!\frac{1}{2}
    \begin{pmatrix}
        e^{-i(A+B)}+e^{-i(A-B)} & e^{-i(A+B)}-e^{-i(A+B)}\\
        e^{-i(A+B)}-e^{-i(A-B)} & e^{-i(A+B)}+e^{-i(A-B)}
    \end{pmatrix}.
\end{align}

A procedure for diagonalizing $3\times3$ matrices is presented in the Appendix.~\ref{Appendix:new_Fourier3}. After some algebra, the resulting matrix elements for $U=e^{-i\mathcal{C}}$ for the coupling matrix in Eq.~\eqref{eq_appe:C_6x6} are given by
\begin{subequations}\label{eq_appe:U_6x6}
    \begin{align}
    \nonumber U_{1,1} &= \frac{1}{2}\left(\cos{(\theta_2)} +\cos{\left(\sqrt{2}\theta_1\right)}\cos{\left(\sqrt{2}\varphi\right)}\cos{\left(\theta_2\right)} \right),\\
    &+\frac{i}{2}\sin{\left(\sqrt{2}\theta_1\right)}\sin{\left(\sqrt{2}
    \varphi\right)}\sin{\left(\theta_2\right)},\\
    \nonumber U_{1,2} &= -\frac{1}{2}\sin{\left(\sqrt{2}\theta_1\right)}\sin{\left(\sqrt{2}\varphi\right)}\cos{\left(\theta_2\right)}\\
    &-\frac{i}{2}\left(1 +\cos{\left(\sqrt{2}\theta_1\right)}\cos{\left(\sqrt{2}\varphi\right)}\right)\sin{(\theta_2)},\\
    U_{1,3} \!&= \!\frac{-ie^{-i\theta_2}}{2\sqrt{2}}\left(e^{2i\theta_2}\sin{\!\left(\!\sqrt{2}(\theta_1-\varphi)\!\right)}+\sin{\!\left(\!\sqrt{2}(\theta_1+\varphi)\!\right)}\!\right),\\
    \nonumber U_{1,4} &= -\frac{1}{2}\sin{\left(\sqrt{2}\theta_1\right)}\sin{\left(\sqrt{2}\varphi\right)}\cos{\left(\theta_2\right)},\\
    &+\frac{i}{2}\left(1- \cos{\left(\sqrt{2}\theta_1\right)}\cos{\left(\sqrt{2}\varphi\right)}\right)\sin{(\theta_2)},\\
    \nonumber U_{1,5} &= \frac{1}{2}\left(-1+ \cos{\left(\sqrt{2}\theta_1\right)}\cos{\left(\sqrt{2}\varphi\right)}\right)\cos{(\theta_2)},\\
    &+\frac{i}{2}\sin{\left(\sqrt{2}\theta_1\right)}\sin{\left(\sqrt{2}\varphi\right)}\sin{\left(\theta_2\right)}\\
    U_{1,6} &= \frac{ie^{-i\theta_2}}{2\sqrt{2}}\left(e^{2i\theta_2}\sin{\left(\sqrt{2}(\theta_1-\varphi)\right)}-\sin{\left(\sqrt{2}(\theta_1+\varphi)\right)}\right),\\
    \nonumber U_{3,6} &= -\sin{\left(\sqrt{2}\theta_1\right)}\sin{\left(\sqrt{2}\varphi\right)}\cos{(\theta_2)}\\
    &- i\cos{\left(\sqrt{2}\theta_1\right)} \cos{\left(\sqrt{2}\varphi\right)} \sin{(\theta_2)},
\end{align}
\end{subequations}
and the other matrix elements follow the equalities $U_{1,2}=U_{4,5}$, $U_{1,3}=U_{2,6}=U_{3,5}=U_{4,6}$, $U_{1,4}=U_{2,5}$
$U_{1,5}=U_{2,4}$, $U_{1,6}=U_{2,3}=U_{3,4}=U_{5,6}$.

First, we have to find the couplings to meet the normalization condition $|U_{jk}|^2=1/6$. However, as can be clearly appreciated in Eqs.\eqref{eq_appe:U_6x6}, we are dealing with an overdetermined system of equations. As there are three variables, the couplings $\theta_1$, $\theta_2$ and $\varphi$, we select the matrix elements $U_{1,2}$, $U_{1,4}$ and $U_{3,6}$ to calculate the square modulus, that is,
\begin{subequations}
    \begin{align}
    \nonumber \lvert U_{1,2}\rvert^2 &=  \frac{1}{4}\sin^2{\left(\sqrt{2}\theta_1\right)}\sin^2{\left(\sqrt{2}\varphi\right)}\cos^2{(\theta_2)}\\
    \nonumber &+\frac{\sin^2{(\theta_2)}}{4} + \frac{\sin^2{(\theta_2)}}{2}\cos{\left(\sqrt{2}\theta_1\right)}\cos{\left(\sqrt{2}\varphi\right)}\\
    \label{eq_appe:U12_6x6}&+\frac{\sin^2{(\theta_2)}}{4}\cos{\left(\sqrt{2}\theta_1\right)}\cos{\left(\sqrt{2}\varphi\right)},\\
    \nonumber \lvert U_{1,4}\rvert^2 &= \frac{1}{4}\sin^2{\left(\sqrt{2}\theta_1\right)}\sin^2{\left(\sqrt{2}\varphi\right)}\cos^2{(\theta_2)}\\
    \nonumber &+\frac{\sin^2{(\theta_2)}}{4} - \frac{\sin^2{(\theta_2)}}{2}\cos{\left(\sqrt{2}\theta_1\right)}\cos{\left(\sqrt{2}\varphi\right)}\\
    \label{eq_appe:U14_6x6}&+\frac{\sin^2{(\theta_2)}}{4}\cos{\left(\sqrt{2}\theta_1\right)}\cos{\left(\sqrt{2}\varphi\right)},\\
    \nonumber \lvert U_{3,6}\lvert^2 &= \sin^2{\left(\sqrt{2}\theta_1\right)}\sin^2{\left(\sqrt{2}\varphi\right)}\cos^2{(\theta_2)}\\
    \label{eq_appe:U36_6x6}&+\cos^2{\left(\sqrt{2}\theta_1\right)}\cos^2{\left(\sqrt{2}\varphi\right)}\sin^2{(\theta_2)}.
\end{align}
\end{subequations}

Using the value $1/6$ for the square modulus of Eq.~\eqref{eq_appe:U12_6x6} and Eq.~\eqref{eq_appe:U14_6x6}, after subtracting them, we obtain
\begin{align}\label{eq_appe:theta_1}
    0 &=\cos{\left(\sqrt{2}\theta_1\right)}\cos{\left(\sqrt{2}\varphi\right)},\\
    \theta_1 &= \pm \frac{\pi}{2\sqrt{2}}+k_1\frac{\pi}{\sqrt{2}}.
\end{align}

Using the solution we found for $\theta_1$ in Eq.~\eqref{eq_appe:U36_6x6}, we arrive at
\begin{align}\label{eq_appe:theta2_phi}
    \frac{1}{6} &=\sin^2{\left(\sqrt{2}\varphi\right)}\cos^2{(\theta_2)}.
\end{align}

Substituting the solution for $\theta_1$~\eqref{eq_appe:theta_1} in Eq.~\eqref{eq_appe:U12_6x6}, we obtain
\begin{align}\label{eq_appe:phi_theta2}
    \frac{1}{6} &= \frac{1}{4}\sin^2{\left(\sqrt{2}\varphi\right)}\cos^2{\left(\theta_2\right)}+\frac{1}{4}\sin^2{\left(\theta_2\right)}.
\end{align}
Using the whole expression of Eq.~\eqref{eq_appe:theta2_phi} in Eq.~\eqref{eq_appe:phi_theta2}, we have the following result
\begin{align}
    \nonumber \frac{1}{6} &= \frac{1}{4}\frac{1}{6}+\frac{1}{4}\sin^2{\left(\theta_2\right)},\\
    \nonumber \sin{\left(\theta_2\right)} &= \pm \frac{\sqrt{2}}{2},\\
    \theta_2 &= \pm\frac{\pi}{4}+k_2\pi.
\end{align}

Using the solution for $\theta_2$ in Eq.~\eqref{eq_appe:theta2_phi}, we finally obtain $\varphi$
\begin{align}
    \nonumber \frac{1}{6} &=\sin^2{\left(\sqrt{2}\varphi\right)}\frac{1}{2},\\
    \varphi &=\pm \frac{1}{\sqrt{2}}\arcsin{\left(1/\sqrt{3}\right)}+k_0\frac{\pi}{\sqrt{2}}
\end{align}

The matching of arguments with the Fourier matrix can be easily checked by replacing the solutions for $\theta_1$, $\theta_2$ and $\varphi$ in Eqs.~\eqref{eq_appe:U_6x6}. Note that $U$ is not dephased.

\section{New 3-dimensional QFTI\label{Appendix:new_Fourier3}}

Adding a different propagation constant to the central waveguide in the nearest-neighbor line, the coupling matrix reads
\begin{align}\label{eq_appe:3x3}
    \mathcal{C} &= 
    \begin{pmatrix}
        0 & \theta & 0\\
        \theta & \beta & \theta\\
        0 & \theta & 0
    \end{pmatrix}.
\end{align}

To calculate the exponential matrix of $\mathcal{C}$ we diagonalized it
\begin{align}\label{eq_appe:exp_C_3x3}
    e^{-i\mathcal{C}} &= Ve^{-i\Lambda}V^\dagger,
\end{align}
where $V$ is the matrix of eigenvectors and $\Lambda$ is a diagonal matrix with eigenvalues of $\mathcal{C}$. In this case, the characteristic polynomial of Eq.~\eqref{eq_appe:3x3} is given by
\begin{align}
    p(\lambda)= -\lambda(\lambda^2-\beta\lambda-2\theta^2),
\end{align}
and consequently, the eigenvalues are
\begin{subequations}\label{eq_appe:eigenvalues_3x3}
    \begin{align}
        \lambda_1 &=0,\\
        \lambda_2 &=\frac{\beta}{2}-\frac{1}{2}\sqrt{\beta^2+8\theta^2},\\
        \lambda_3 &=\frac{\beta}{2}+\frac{1}{2}\sqrt{\beta^2+8\theta^2}.
    \end{align}
\end{subequations}
On the other hand, the normalized eigenvectors are
\begin{subequations}\label{eq_appe:eigenvectors_3x3}
    \begin{align}
    \vec v_1 &= \frac{1}{\sqrt{2}}
    \begin{pmatrix}
        -1\\
        0\\
        1
    \end{pmatrix},\\
    \vec v_2 &= \frac{1}{\sqrt{2\Theta^2-2\beta\Theta}}
    \begin{pmatrix}
        2\theta\\
        \beta-\Theta\\
        2\theta
    \end{pmatrix},\\
    \vec v_3 &=\frac{1}{\sqrt{2\Theta^2+2\beta\Theta} }
    \begin{pmatrix}
        2\theta\\
        \beta+\Theta\\
        2\theta
    \end{pmatrix},
\end{align}
\end{subequations}
where $\Theta^2=\beta^2+8\theta^2$.

For this case, let us denote the transformation matrix as $U=U^{(3)}$. Putting together the eigenvalues~\eqref{eq_appe:eigenvalues_3x3} and eigenvectors~\eqref{eq_appe:eigenvectors_3x3}
in Eq.~\eqref{eq_appe:exp_C_3x3}, the elements of $U^{(3)}$ are given by
\begin{subequations}\label{eq_appe:U_3x3}
    \begin{align}
    U_{1,1}^{(3)} &= \frac{1}{2}+\frac{e^{-i\beta/2}}{2\Theta} \left(\Theta\cos{\left(\Theta/2\right)}+i\beta\sin{\left(\Theta /2\right)}\right),\\
    U_{1,2}^{(3)} &= -\frac{2i\theta}{\Theta}e^{-i\beta/2}\sin{\left(\Theta/2\right)},\\
    U_{1,3}^{(3)} &= -\frac{1}{2}+\frac{e^{-i\beta/2}}{2\Theta} \left(\Theta\cos{\left(\Theta/2\right)}+i\beta\sin{\left(\Theta /2\right)}\right),\\
    U_{2,2}^{(3)} &= e^{-i\beta /2}\left(\cos{\left(\Theta/2\right)}-i\frac{\beta }{\Theta}\sin{\left(\Theta/2\right)}\right),\\
    \Theta &= \sqrt{\beta^2+8\theta^2},
\end{align}
\end{subequations}
where $U_{33}^{(3)}=U_{11}^{(3)}$ and $U_{23}^{(3)}=U_{12}^{(3)}$.

First, we have to check the normalization condition $|U_{jk}|^2=1/N$. For this purpose, we need the square moduli of Eqs.~\eqref{eq_appe:U_3x3}, that is
\begin{align}
    \nonumber \left|U_{1,1}^{(3)}\right|^2 &=\frac{1}{4}+\frac{1}{2}\cos{(\beta/2)}\cos{(\Theta/2)}\\
    \nonumber &\hphantom{=}+\frac{1}{2\Theta}\beta\sin{(\beta/2)}\sin{(\Theta/2)}\\
    \label{eq_appe:U3_11}&\hphantom{=}+\frac{1}{4\Theta^2}\left(\Theta^2\cos^2{(\Theta/2)}+\beta^2\sin^2{(\Theta/2)}\right),\\
    \label{eq_appe:U3_12}\left|U_{1,2}^{(3)}\right|^2 &= \frac{4\theta^2}{\Theta^2}\sin^2{\left(\Theta/2\right)},\\
    \nonumber \left|U_{1,3}\right|^2 &=\frac{1}{4}-\frac{1}{2}\cos{(\beta/2)}\cos{(\Theta/2)}\\
    \nonumber &\hphantom{=}-\frac{1}{2\Theta}\beta\sin{(\beta/2)}\sin{(\Theta/2)}\\
    \label{eq_appe:U3_13}&\hphantom{=}+\frac{1}{4\Theta^2}\left(\Theta^2\cos^2{(\Theta/2)}+\beta^2\sin^2{(\Theta/2)}\right).
\end{align}

Let us assume that a solution for $\beta$ and $\theta$ exist and $|U^{(3)}_{1,1}|^2=|U_{1,3}^{(3)}|^2=1/3$, this implies that by summing both equations, we obtain
\begin{align}
    \frac{2}{3} &= \frac{1}{2}+\frac{1}{2\Theta^2}\left(\Theta^2\cos^2{\left(\Theta/2\right)}+\beta^2\sin^2{\left(\Theta/2\right)}\right),\\
    \frac{1}{6} &= \frac{1}{2\Theta^2}\left(\beta^2\cos^2{\left(\Theta/2\right)}+8\theta^2\cos^2{\left(\Theta/2\right)}+\beta^2\sin^2{\left(\Theta/2\right)} \right),\\
    \frac{1}{6} &= \frac{1}{2\Theta^2}\left(\Theta^2-8\theta^2\sin^2{\left(\Theta/2\right)} \right),\\
    \frac{1}{3} &= \frac{4\theta^2}{\Theta^2} \sin^2{\left(\Theta/2\right)}.
\end{align}
This is precisely the normalization condition for $|U_{12}|^2$, so the assumption is consistent and shows that $U^{(3)}$ can be balanced. All complex Hadamard matrices of dimension $3$ are equivalent to the Fourier matrix\cite{Tadej2006}, so we do not need to check the arguments.

\bibliography{ref}

\bibliographystyle{apsrev4-2}

\end{document}